\documentclass[twocolumn]{aastex62}

\usepackage{savesym}
\savesymbol{tablenum}
\usepackage{siunitx}
\usepackage{amsmath}

\newlength{\apjcolwidth}
\newlength{\figwidth}
\shorttitle{Pulsational Pair-Instability Supernovae in very close binaries}
\shortauthors{Marchant et al.}

\begin{document}

\title{Pulsational pair-instability supernovae in very close binaries}

\correspondingauthor{Pablo Marchant}
\email{pablo.marchant@northwestern.edu}

\author[0000-0002-0338-8181]{Pablo Marchant}
\affil{Center for Interdisciplinary Exploration and Research in Astrophysics
(CIERA) and Department of Physics and Astronomy, Northwestern University, 2145 Sheridan Road, Evanston, IL 60208, USA}

\author[0000-0002-6718-9472]{Mathieu Renzo}
\affiliation{Anton Pannenkoek Institute for Astronomy, University of Amsterdam,
NL-1090 GE Amsterdam, the Netherlands}
\affiliation{GRAPPA, University of Amsterdam, Science Park 904, 1098 XH Amsterdam, The Netherlands}

\author[0000-0003-3441-7624]{Robert Farmer}
\affiliation{Anton Pannenkoek Institute for Astronomy, University of Amsterdam,
NL-1090 GE Amsterdam, the Netherlands}
\affiliation{GRAPPA, University of Amsterdam, Science Park 904, 1098 XH Amsterdam, The Netherlands}

\author[0000-0003-4250-6766]{Kaliroe M. W. Pappas}
\affiliation{Department of Physics, University of Illinois at Urbana-Champaign, 1110 W Green St Loomis Laboratory, Urbana, IL 61801, USA}

\author[0000-0001-8805-2865]{Ronald E. Taam}
\affil{Center for Interdisciplinary Exploration and Research in Astrophysics
(CIERA) and Department of Physics and Astronomy, Northwestern University, 2145 Sheridan Road, Evanston, IL 60208, USA}

\author[0000-0001-9336-2825]{Selma E. de Mink}
\affiliation{Anton Pannenkoek Institute for Astronomy, University of Amsterdam,
NL-1090 GE Amsterdam, the Netherlands}
\affiliation{GRAPPA, University of Amsterdam, Science Park 904, 1098 XH Amsterdam, The Netherlands}

\author[0000-0001-9236-5469]{Vassiliki Kalogera}
\affil{Center for Interdisciplinary Exploration and Research in Astrophysics
(CIERA) and Department of Physics and Astronomy, Northwestern University, 2145 Sheridan Road, Evanston, IL 60208, USA}

%
%
%
%



\begin{abstract}
   Pair-instability and pulsational pair-instability supernovae (PPISN) have not been unambiguously
   observed so far. They are, however, promising candidates for the progenitors of the heaviest
   binary black hole (BBH) mergers detected. If these
   BBHs are the product of binary evolution, then PPISNe could occur in
   very close binaries.
   Motivated by this, we discuss the implications of a PPISN happening with a
   close binary companion, and what impact these events have on the
   formation of merging BBHs through binary evolution. For this, we have computed a set of
   models of metal-poor ($Z_\odot/10$) single helium stars using the
   \texttt{MESA} software instrument.
   For PPISN progenitors with pre-pulse masses $>50M_\odot$ we find that, after a pulse,
   heat deposited throughout the layers of the star that
   remain bound cause it to expand to more than $100R_\odot$ for periods of
   $10^2-10^4\;$~yrs depending on the  mass of the progenitor. This results in long-lived phases of Roche-lobe overflow
   or even common-envelope events if there is a close binary companion, leading
   to additional electromagnetic transients associated to PPISN eruptions.
   If we ignore the effect of these interactions, we find that mass loss from PPISNe reduces the
   final black hole spin by $\sim 30\%$, induces eccentricities 
   below the threshold of detectability of the LISA observatory, and can produce a
   double-peaked distribution of measured chirp masses in BBH mergers observed
   by ground-based detectors.
\end{abstract}



\section{Introduction} \label{sec:intro}
The production of electron-positron pairs in the cores of very massive stars has long been proposed to
cause their collapse before oxygen is depleted in their cores, leading to a
thermonuclear explosion \citep{FowlerHoyle1964,RakaviShaviv1967}. Stars with helium cores between $M_{\rm He}\sim
60-130 M_\odot$ (corresponding to zero-age main-sequence masses between $\sim
140-260M_\odot$ for non-rotating stars without mass-loss) are expected to be completely disrupted by this event, with the
higher mass progenitors possibly being observable as super-luminous supernovae (SNe) owing to nickel
yields of up to tens of solar masses \citep{HegerWoosley2002}. Less massive stars, with helium
core masses in the range of $M_{\rm He}\sim 30-60 M_\odot$ (zero-age
main-sequence masses $\sim 70-140M_\odot$), are also expected
to become unstable but produce instead a series of energetic pulses and mass
ejections before finally collapsing to a black hole (BH,
\citealt{Fraley1968,Woosley2017}). These two types of events are referred to as pair-instability
supernovae (PISN) and pulsational pair-instability supernovae (PPISN)
respectively. Stars with cores in excess of $M_{\rm He}\sim 130 M_\odot$ are also predicted to become unstable,
but energy losses due to photodisintegration of heavy elements
prevent a thermonuclear explosion and allow the formation of a BH
\citep{WoosleyWeaver1982,Bond+1982, HegerWoosley2002}.
Although no observed SN has been conclusively identified to be either a PISN or
a PPISN, theoretical models consistently predict these transients, with
physical uncertainties such as rotation \citep{ChatzopoulosWheeler2012} or
nuclear reaction rates \citep{Takahashi2018} only shifting the mass ranges listed above.

Various potential candidate events from hydrogen-rich SNe have been observed.
OGLE-2014-SN-073 is one such PISN candidate, with a derived ejecta mass of
$60^{+45}_{-16}M_\odot$ and a nickel mass $>0.47\pm0.02M_\odot$ \citep{Terreran+2017}.
SN 2006gy on the other hand has been proposed to be powered by the collision of
ejected shells in a PPISN \citep{Woosley+2007}. This is also the case for
iPTF14hls, as its light curve
exhibits multiple peaks and a high brightness for more than $600\;\rm days$
(\citealt{Arcavi+2017}, see \citealt{Woosley2018} for a discussion on potential
progenitors). Regarding hydrogen-poor events, the type I superluminous SNe
SN 2007bi has been suggested
to be the product of a PISN with a nickel yield $>3M_\odot$
\citep{Gal-Yam+2009}.
Another type I superluminous event, iPTF16eh
produced a light-echo on a shell of material ejected $\sim32$ years
prior to explosion \citep{Lunnan+2018}, making it a prime candidate for a PPISN.
Upcoming transient surveys such as the ZTF
\citep{Bellm2014,Smith+2014} and the
LSST \citep{Abell+2009} will detect similar events in large numbers, providing
vital information to establish or discard their origin as pair-instability driven
transients (although note that the light echo of iPTF16eh was detected through
flash-spectroscopy, and would be missed by photometric surveys).

In this context, the detection of merging binary BHs (BBHs) by the advanced LIGO
(aLIGO) and Virgo (aVirgo) detectors \citep{Abbott_LIGOsummary_2016} can provide indirect evidence of the
existence of PISNe and PPISNe. If these sources are formed via stellar binary
evolution in the field, PISN are expected to produce a clear gap in the observed
masses of merging BBHs \citep{Belczynski+2014, Marchant+2016}. PPISN are
expected to widen this gap, as BH progenitors just below the PISN threshold can
lose more than $10M_\odot$ before collapse \citep{Woosley2017}. Given the
sensitivity of the aLIGO detectors and the reported BBH mergers at
the time, \citet{FishbachHolz2017} showed there
was an indication of an upper mass cutoff of $\sim 40M_\odot$,
consistent with models of field binary evolution which
include both PISN and PPISN \citep{Belczynski+2016b,SperaMapelli2017}.
The recent release of the first Gravitational Wave Transient Catalog
by the LIGO-Virgo collaboration includes a total of 10 BBH merger
detections \cite{GWTC1}, and strongly favours a dearth of BH masses above
$45M_\odot$ \cite{LIGOpop}.

Theoretical work to explain the formation of merging compact
objects was driven at first by the discovery of the Hulse-Taylor binary pulsar
\citep{HulseTaylor1975}. Common envelope (CE) evolution, which had been proposed
by \citet{Paczynski1976} as the formation mechanism of cataclysmic variables,
was invoked by \citet{vandenHeuvel1976} to reduce the orbital separations of
wide massive binaries and produce close binary neutron stars. As it was realized
that the Hulse-Taylor pulsar would coalesce due to gravitational wave (GW)
emission, population synthesis studies were done to understand the rate of such
events in the context of CE evolution \citep{Clark1979}. More than a decade
later, CE evolution was also proposed to form merging BBHs
\citep{TutukovYungelson1993}, and that BBH mergers would be the dominant sources
detectable by ground-based GW observatories (\citealt{Lipunov1997}, see
\citealt{Dominik+2012} and \citealt{Belczynski+2016} for more recent work).

Various additional channels have been put forward to explain the origin of
merging BBHs. In very close binaries, efficient rotational mixing has been
predicted to lead to merging BBHs, as
chemically homogeneous evolution (CHE)
prevents the expansion of a star during its main sequence
\citep{Maeder1987} and allows for an
initially compact binary to remain so until BH formation
\citep{MandeldeMink2016, Marchant+2016, deMinkMandel2016}.
BBHs can also be formed through dynamical interactions in
dense environments \citep{Kulkarni+1993, SigurdssonHernquist1993}, with large
systems such as globular clusters producing BBHs compact enough to merge
\citep{PortegieszwartMcmillan2000}. Other scenarios include formation in triple
star systems (c.f. \citealt{Thompson2011,Antonini+2014,Antonini+2017}) and AGN disks \citep{Bartos+2017,Stone+2017}.

The objective of this work is to study the implications of PPISNe occurring with
nearby binary companions, and what effects the pulses have
on the resulting BHs that could be observed through the detection of
GWs in BBH mergers.
To do this, we perform detailed
simulations of the formation of BHs from single helium stars undergoing PPISN.
These are appropriate to model BBHs formed through binary evolution,
including the CE and CHE channels, where
each star is expected to become hydrogen poor at its surface before BH
formation. In Section
\ref{sec:methods} we describe our methods and present our PPISN models in
Section \ref{sec:results}.
In Section \ref{sec:ppisn2} we
discuss how the presence of a nearby companion can affect the occurrence of a
PPISN while in Section \ref{sec:mergingBBH} we describe how PPISNe affect the
observable properties of a merging BBH.
We conclude with a discussion of our results in Section
\ref{sec:discussion}.
All our models are available for download at
\url{https://doi.org/10.5281/zenodo.1211427}, including the input
files to perform these simulations, machine readable tables, and movies for each
of our simulations.

\section{Methods} \label{sec:methods}
We compute a set of
non-rotating models of helium stars at a metallicity of $Z_\odot/10$, defining
$Z_\odot=0.0142$ as the proto-solar abundance reported by \citet{Asplund+2009}.
Our simulations are computed using version 11123 of the \texttt{MESA} software instrument for stellar evolution
\citep{Paxton+2011,Paxton+2013,Paxton+2015,Paxton+2018}.
Radiative opacities are computed using tables from the OPAL
project \citep{IglesiasRogers1996}.
Convective regions are
determined using the Ledoux criterion and convective energy transport and mixing is modelled
using a prescription for time-dependent convection which we describe in Appendix
\ref{appendix:tdc}. Regions that are stable according to the Ledoux criterion but unstable
according to the Schwarzschild criterion undergo semiconvective mixing, which
we model following \cite{Langer+1983} with an efficiency parameter of
$\alpha_{\rm sc}=1$. Overshooting from convective boundaries is modelled using
exponential overshooting \citep{Herwig2000} with a parameter $f=0.01$.
Note that, formally, convective velocities are zero at the edge
of a convective zone, such that an additional parameter $f_0$ is required to
define expotential overshooting. The evaluation of the exponentially decaying
mixing coefficient is then done at a distance $f_0H_P$ inside the convective
boundary, and we choose a value of $f_0=0.005$.
Our chosen treatment softens convective boundaries and allow convective
regions to expand against steep composition gradients.
As a reference,
\citep{Herwig2000} finds that $f=0.016$ is required for convective hydrogen
burning cores to reproduce the width of
the main sequence.

Nuclear reactions are computed using the
\texttt{basic}, \texttt{co\_burn} and \texttt{approx21} nuclear networks
provided in \texttt{MESA} which are switched during runtime to account for
different phases of nuclear burning. In particular, during pulsational phases we
use the \texttt{approx21} network. We provide a detailed description of this
21-isotope network and discuss how appropriate it is for these evolutionary phases in Appendix
\ref{app:conv}, where we also present the results of a convergence test using a larger network. Nuclear
reaction rates are taken from \citet{CaughlanFowler1988} and \citet{Angulo+1999}
with preference given to the latter when available.

Our modelling of stellar winds follows that of \citet{Brott+2011}. All our
models are hydrogen depleted at their surface, so we adopt the mass loss rates
of \citet{Hamann+1995}, scaled by a factor of $1/10$ to account for the effect
of clumping \citep{Yoon+2010}. Although we only model naked helium stars, as
we will show in Section \ref{sec:ppisn2} energy deposited by a
PPISNe on the outer layers can make the envelope of these stars expand requiring
also a recipe for winds from cool stars. For this purpose we take the mass loss
rate to be the maximum between the mass loss rates of
\citet{NieuwenhuijzendeJager1990} and a tenth of the rate from \citet{Hamann+1995}.
The rates provided by \citet{NieuwenhuijzendeJager1990} are calibrated using
stars on our galaxy, so to account for lower mass loss rates at lower
metallicities, we scale it by a factor of $(Z/Z_\odot)^{0.85}$. This assumes the
scaling of cool winds with metallicity follows the dependence for hot stars
derived by \citet{Vink+2001}, which is consistent with observations of
OB stars in the Galaxy and the Magellanic clouds \citep{Mokiem+2007b}.

\subsection{Modelling of PPISNe} \label{sec:modppi}
Up to central helium depletion we assume hydrostatic equilibrium in our models.
At later phases, we consider the weighted value of the first adiabatic
exponent,
\begin{eqnarray}
   \langle \Gamma_1\rangle = \frac{\int_0^M \frac{\Gamma_1P}{\rho} dm}
   {\int_0^M \frac{P}{\rho} dm},\qquad
   \Gamma_1=\left(\frac{\partial \ln P}{\partial \ln \rho}\right)_{\rm ad}
   \label{eq:gamma}
\end{eqnarray}
where $M$ is the total mass of the star. The condition
$\langle\Gamma_1\rangle<4/3$ can then be used as an approximate stability
criterion (cf. \citealt{Stothers1999}) to determine when the assumption of
hydrostatic equilibrium is inappropriate. In our simulations, whenever
$\langle\Gamma_1\rangle-4/3 < 0.01$ and the central temperature exceeds $10^9$
K, instead of assuming hydrostatic equilibrium we use the HLLC solver for hydrodynamics 
\citep{Toro+1994} which has recently been implemented into
\texttt{MESA} \citep{Paxton+2018}. This method can accurately model shocks and
preserve energy, without requiring the use of an artificial viscosity.
To account for iron-core collapse or rapid
evolution due to neutrino emission before the onset of dynamical instability,
we also switch to the HLLC solver if the central temperature exceeds
$10^{9.6}$ K
or the neutrino luminosity is above $10^{10}L_\odot$. Wind mass loss is ignored when the HLLC
solver is in use. For models that result in PPISN and PISN, we define the
first instance when $\langle\Gamma_1\rangle-4/3 < 0.005$ as the pre-supernova
stage.

Modelling PPISNe is
particularly challenging, as after a pulse the star can settle back into
hydrostatic equilibrium and undergo periods of quiescence of more than a thousand
years \citep{Woosley2017}. As the ejected layers expand and cool
down, they become optically thin and go outside the range of applicability of
\texttt{MESA}. To avoid this, during these long inter-pulse periods we remove
the unbound layers as described in Appendix \ref{app:relax} and switch to a
hydrostatic model if the conditions to turn on hydrodynamics described above are
not met.

In order to distinguish individual pulses from our models, we compute at each
step the maximum velocity in the inner $95\%$ of the star that remains below the
local escape velocity $v_{\rm esc}=\sqrt{2Gm/r}$, which we define as $v_{95}$.
Whenever $|v_{95}|>20\;\rm km\;s^{-1}$ we consider that instant to be the beginning of a
pulsation. After this point, we consider a pulse to finish once the inner layers
are close to hydrostatic equilibrium. To do this, we take into account a dynamical
timescale $\tau_{95}=1/\sqrt{G\langle\rho\rangle_{95}}$, where
$\langle\rho\rangle_{95}$ is the average density of the inner $95\%$ of mass
that remains bound. Whenever $|v_{95}|<20\;\rm km\;s^{-1}$ for a time longer than
$20\tau_{95}$, or if the star undergoes iron-core collapse,
we consider the pulse is finished. Even if the conditions for our definition of
a pulse are met, we discard it if it results in ejections of less than
$10^{-6}M_\odot$, which also prevents iron-core collapse from being defined as a
pulse. Although the values chosen are arbitrary, we
have verified for all models computed that they match a by-eye definition of
each mass ejection. Having a well defined criterion gives us a way to
unambiguously identify each pulsation.

Except for cases that are near the limit between PPISN and PISN, all our models
that undergo iron-core collapse have final masses in excess of $20M_\odot$. For
such large core masses we expect a BH to be formed through direct collapse
\citep{Fryer1999} and for all our models we assume the final BH mass $M_{\rm BH}$ is equal to the
baryonic mass before iron-core collapse. Note however that some recent
simulations have resulted in BH formation through fallback in
a succesful explosion, instead of direct
collapse, even for such massive helium cores \citep{Chan+2018, Ott+2018, Kuroda+2018}. This
would further reduce the mass of the final remnant.

\section{Single star models} \label{sec:results}
Before discussing the overall properties of our models, we show the evolution of
two representative PPISNe simulations corresponding to helium stars with initial
masses of $46M_\odot$ and $76M_\odot$, and compare their mass loss
and kinetic energy of ejecta to the
models of \citet{Woosley2017} which are computed at zero-metallicity and without
winds.

\begin{figure}
   \includegraphics[width=\columnwidth]{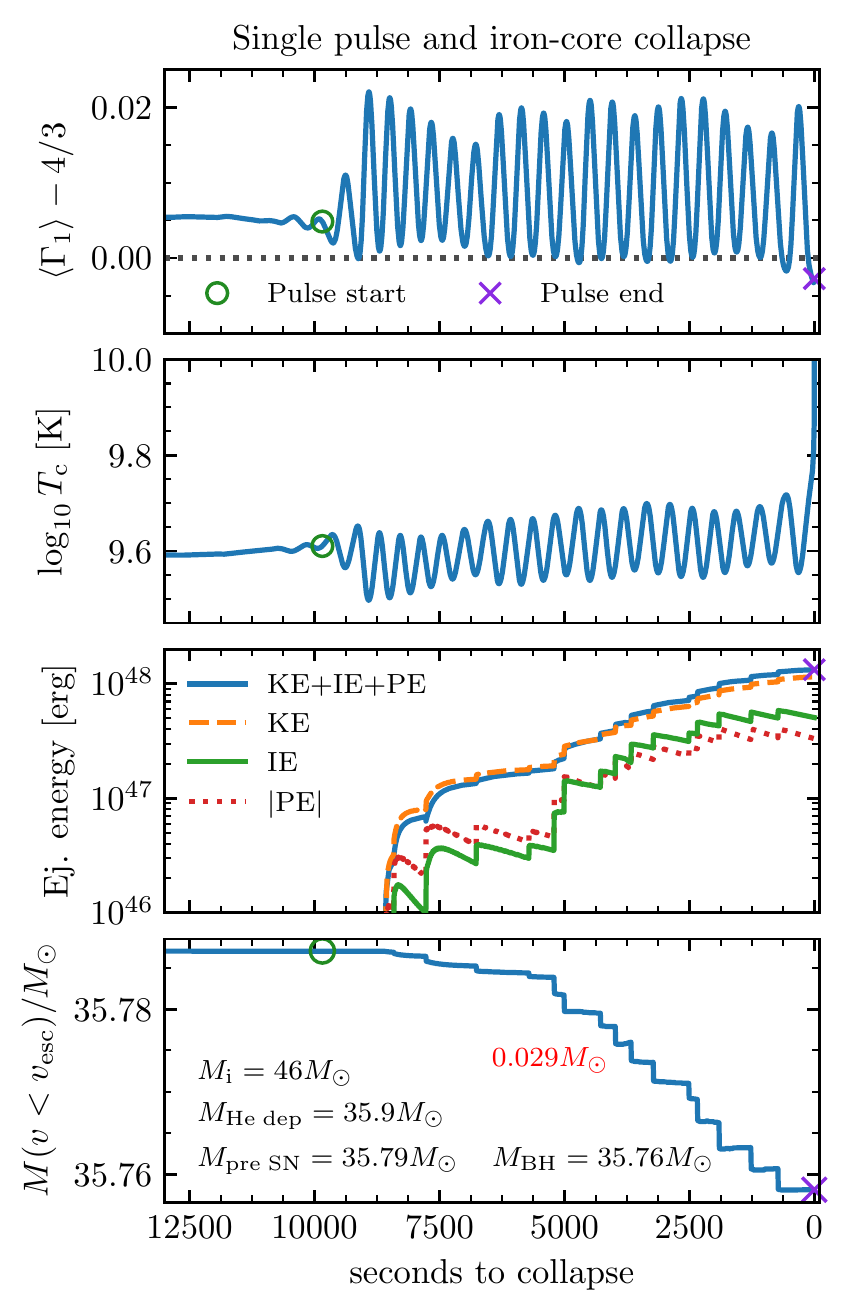}
   \caption{
      Evolution of a PPISN from a helium star model with an initial mass
   $M_i=46M_\odot$. Panels show the evolution with time of the average 
   $\langle\Gamma_1\rangle$ (see Eq. (\ref{eq:gamma})), the central temperature, the energy of layers
   ejected due to the pulse and the total mass of the star that remains below
   the escape velocity. The energy of ejected layers is also separated into its
   kinetic energy (KE), internal energy (IE) and potential energy (PE).
   The mass ejected through the pulse is written down in red in
   the bottom panel. Symbols are also used to denote the beginning and the end
   of the pulse as defined in Section \ref{sec:modppi}.
   \label{fig:pulses46}}
\end{figure}

The $46M_\odot$ model reaches core helium depletion with a mass of $M_{\rm He
\;dep}=35.9M_\odot$. It then undergoes
hydrostatic core-carbon burning and hydrostatic
core-oxygen burning. Only after oxygen in
the core has been depleted, the star contracts into the
pair-creation region, leading to a reduced $\langle\Gamma_1\rangle$.
As the star approaches $\langle
\Gamma_1\rangle=4/3$ it starts experiencing oscillations, and the burning of
carbon and oxygen in shells provide sufficient energy to eject $0.0289M_\odot$
with a kinetic energy of $1.2\times 10^{48}$ erg, as shown in Figure
\ref{fig:pulses46}. Only $3$ hours pass from the onset of the
instability to iron-core collapse and the star never recovers hydrostatic
equilibrium during this time. According to our definition in the previous
section we then consider this to be an individual pulse\footnote{Note that this
definition is different from the PPISN calculations of \citet{Woosley2017}, who
labels each individual oscillation in such cases as a 'weak pulse'.}, after which a
$35.8M_\odot$ BH is formed.

\begin{figure*}[ht!]
   \includegraphics[width=\textwidth]{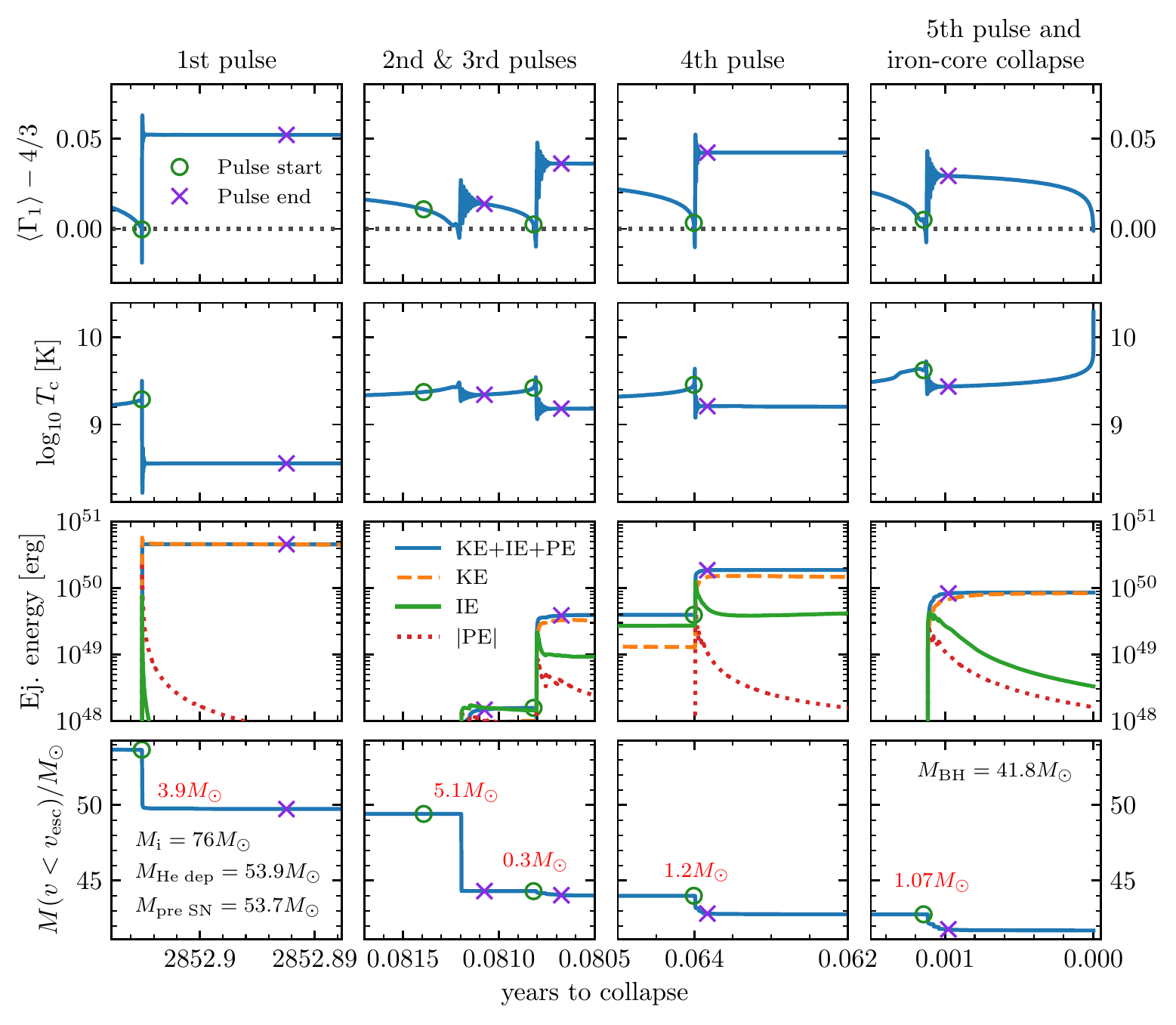}
   \caption{
    Same as Figure \ref{fig:pulses46} but for a PPISN model with an
   initial mass $M_{i}=76M_\odot$. This star however undergoes a single phase of
   dynamical instability before iron-core collapse.
   Each set of vertical
   panels shows a zoomed-in time evolution during different pulses.
   \label{fig:pulses76}}
\end{figure*}

The $76M_\odot$ model also results in a PPISN, but its evolution is dramatically
different. This simulation reaches core-helium depletion
with its mass lowered to $M_{\rm He\;dep}=53.9M_\odot$ due to stellar winds,
after which it undergoes hydrostatic core-carbon burning.
Pair-creation then leads to a reduced $\langle\Gamma_1\rangle$ and dynamical
instability before core-oxygen ignition, and we depict its pulsational stage in
Figure \ref{fig:pulses76}.
When $\langle\Gamma_1\rangle-4/3<0.005$, which is the point we have
defined as the pre-SN stage, winds have further reduced the mass of the star
by $0.2M_\odot$ down to $M_{\rm pre\;SN}=53.7M_\odot$.
At this moment it experiences a strong pulsation that removes
$3.94M_\odot$ with a kinetic energy of $5\times10^{50}$ ergs and lowers the
central temperature by almost a factor of 4 compared to its value at the
beginning of the pulse. The star then has a long quiescence phase
lasting almost $3000$ years, until it again becomes pulsationally unstable,
leading to additional pulses and mass loss within a month of iron-core collapse.
Although a pulse happens just three days prior to iron-core collapse, the star
returns to equilibrium and undergoes hydrostatic core-silicon burning before
collapsing to a $41.8M_\odot$ BH.

\begin{deluxetable*}{cccccccccc}[ht!]
\tablecaption{Hydrogen free PPISN models at low metallicity ($Z=Z_\odot/10$)\label{tab:models}}
\tablecolumns{9}
\tablenum{1}
\tablewidth{0pt}
\tablehead{
\colhead{$M_{\rm initial}$} &
\colhead{$M_{\rm He\;dep}$} &
\colhead{$M_{\rm CO}$} & 
\colhead{$M_{\rm pre\;SN}$} &
\colhead{$M_{\rm He,\;pre\;SN}$} &
\colhead{$M_{\rm ejecta}$} & 
\colhead{$M_{\rm BH}$} &
\colhead{\# of pulses} &
\colhead{Duration} &
\colhead{max KE} \\
\colhead{($M_\odot$)} &
\colhead{($M_\odot$)} &
\colhead{($M_\odot$)} &
\colhead{($M_\odot$)} &
\colhead{($M_\odot$)} &
\colhead{($M_\odot$)} &
\colhead{($M_\odot$)} &
\colhead{} &
   \colhead{(yr)} & 
\colhead{$10^{51}\;[\rm erg]$}
}
\startdata
40.00 & 31.99 & 27.69 & - & - & - & 31.87 & 0 & - & - \\
42.00 & 33.32 & 28.92 & - & - & - & 33.19 & 0 & - & - \\
44.00 & 34.63 & 30.12 & - & - & - & 34.50 & 0 & - & - \\
44.50 & 34.95 & 30.42 & - & - & - & 34.82 & 0 & - & - \\
45.00 & 35.28 & 30.72 & 35.14 & 0.69 & 0.02 & 35.13 & 1 & \num{0.000139} & \num{0.000785} \\
45.50 & 35.60 & 31.02 & 35.46 & 0.67 & 0.01 & 35.45 & 1 & \num{1.89e-05} & \num{0.000561} \\
46.00 & 35.92 & 31.32 & 35.79 & 0.68 & 0.03 & 35.76 & 1 & \num{0.00029} & \num{0.00115} \\
48.00 & 37.21 & 32.50 & 37.06 & 0.67 & 0.04 & 37.02 & 1 & \num{0.00129} & \num{0.00127} \\
50.00 & 38.47 & 33.67 & 38.32 & 0.65 & 0.24 & 38.08 & 3 & \num{0.00273} & \num{0.0113} \\
52.00 & 39.73 & 34.82 & 39.57 & 0.64 & 0.66 & 38.90 & 4 & \num{0.00677} & \num{0.0292} \\
54.00 & 40.97 & 35.96 & 40.80 & 0.63 & 0.25 & 40.55 & 5 & \num{0.00485} & \num{0.00929} \\
56.00 & 42.20 & 37.08 & 42.02 & 0.63 & 0.33 & 41.69 & 7 & \num{0.00773} & \num{0.0145} \\
58.00 & 43.41 & 38.21 & 43.23 & 0.62 & 1.62 & 41.60 & 9 & \num{0.0303} & \num{0.108} \\
60.00 & 44.62 & 39.32 & 44.42 & 0.61 & 1.72 & 42.70 & 9 & \num{0.0388} & \num{0.128} \\
62.00 & 45.81 & 40.42 & 45.61 & 0.60 & 2.51 & 43.10 & 10 & \num{0.135} & \num{0.0549} \\
64.00 & 47.00 & 41.50 & 46.79 & 0.60 & 4.12 & 42.67 & 6 & \num{0.622} & \num{0.171} \\
66.00 & 48.17 & 42.57 & 47.95 & 0.59 & 4.55 & 43.41 & 9 & \num{1.37} & \num{0.19} \\
68.00 & 49.33 & 43.66 & 49.11 & 0.58 & 5.29 & 43.82 & 10 & \num{11} & \num{0.202} \\
70.00 & 50.49 & 44.70 & 50.26 & 0.58 & 6.31 & 43.94 & 11 & \num{132} & \num{0.163} \\
72.00 & 51.64 & 45.75 & 51.40 & 0.57 & 8.02 & 43.31 & 6 & \num{732} & \num{0.254} \\
74.00 & 52.78 & 46.80 & 52.53 & 0.57 & 9.61 & 42.72 & 4 & \num{1.9e+03} & \num{0.409} \\
76.00 & 53.92 & 47.87 & 53.66 & 0.56 & 11.66 & 41.69 & 5 & \num{2.85e+03} & \num{0.578} \\
78.00 & 55.05 & 48.89 & 54.79 & 0.56 & 14.06 & 40.33 & 5 & \num{3.84e+03} & \num{0.814} \\
80.00 & 56.18 & 49.94 & 55.91 & 0.55 & 16.81 & 38.64 & 5 & \num{4.67e+03} & \num{1.14} \\
82.00 & 57.31 & 50.97 & 57.02 & 0.55 & 19.22 & 37.33 & 5 & \num{5.38e+03} & \num{1.47} \\
84.00 & 58.42 & 52.00 & 58.13 & 0.54 & 23.73 & 33.94 & 6 & \num{6.18e+03} & \num{1.94} \\
86.00 & 59.51 & 53.01 & 59.20 & 0.54 & 28.98 & 29.89 & 9 & \num{7.13e+03} & \num{2.57} \\
87.00 & 60.04 & 53.49 & 59.73 & 0.54 & 31.66 & 27.82 & 2 & \num{7.5e+03} & \num{2.78} \\
88.00 & 60.58 & 54.01 & 60.27 & 0.54 & 41.58 & 18.60 & 2 & \num{9.72e+03} & \num{3.18} \\
88.50 & 60.85 & 54.27 & 60.54 & 0.54 & 45.26 & 15.22 & 1 & \num{1.08e+04} & \num{3.3} \\
88.75 & 60.96 & 54.38 & 60.65 & 0.54 & 46.97 & 13.63 & 1 & \num{1.14e+04} & \num{3.34} \\
89.00 & 61.10 & 54.48 & 60.79 & 0.54 & 49.18 & 11.57 & 1 & \num{1.29e+04} & \num{3.43} \\
89.05 & 61.13 & 54.52 & 60.81 & 0.54 & 60.81 & - & 1 & - & \num{3.5358} \\
90.00 & 61.64 & 54.95 & 61.32 & 0.53 & 61.32 & - & 1 & - & \num{3.7902} \\
100.00 & 66.80 & 59.77 & 66.44 & 0.52 & 66.44 & - & 1 & - & \num{7.8267} \\
150.00 & 89.93 & 81.23 & 89.37 & 0.50 & 89.37 & - & 1 & - & \num{32.069} \\
200.00 & 109.61 & 99.63 & 108.84 & 0.50 & 108.84 & - & 1 & - & \num{57.553} \\
240.00 & 123.37 & 112.59 & 122.43 & 0.50 & 122.43 & - & 1 & - & \num{63.129} \\
242.00 & 124.12 & 113.29 & - & - & - & 123.18 & 0 & - & - \\
250.00 & 126.61 & 115.65 & - & - & - & 125.64 & 0 & - & - \\
290.00 & 138.87 & 127.14 & - & - & - & 137.74 & 0 & - & -
\enddata
\tablecomments{Initial and final properties of helium stars with a metallicity
   of $Z_\odot/10$ undergoing PPISNe. We define helium depletion as the point in
   time where the central helium mass fraction drops below $0.01$. CO core masses are
   defined when the central helium mass fraction drops below
   $10^{-3}$ as the innermost mass coordinate where the helium mass abundance is
   larger than $0.01$. For pulsating models $M_{\rm pre\;SN}$ is the mass at the
   onset of pulsations, defined as the moment when $\langle \Gamma_1\rangle -4/3 <
   0.005$, and $M_{\rm He,\;pre\;SN}$ is the total mass of helium at
   this point.
   Note that the difference between $M_{\rm pre\;SN}$ and $M_{\rm
   BH}$ is not exactly equal to $M_{\rm ejecta}$, despite our assumption of the
   BH mass being equal to the baryonic mass before iron-core collapse. This
   difference is due to wind mass-loss during phases of quiescence after the
   onset of the first pulse.}
\end{deluxetable*}

This large difference on the timescale from the
onset of pulsations to iron-core collapse is due to the neutrino luminosity
decreasing steeply with central temperature \citep{Woosley+2007,Yoshida+2016,
Woosley2017}. As more massive
models experience more energetic pulses, their post-pulse central temperatures
are lower, resulting in photon radiation from the surface becoming the main
energy-loss mechanism instead of neutrino emission. 
For example, in the first pulse of our
$76M_\odot$ model the neutrino luminosity at the pre-SN stage is in excess of
$10^{12}L_\odot$, and at the end of the pulse it has lowered to $\sim
1.5\times10^{4}L_\odot$. The main source of energy loss at this point is just
radiation from its surface, so the star evolves on the $\sim$ thousand year long
Kelvin-Helmholtz timescale.
Matching the pre-SN mass of our models to the
initial masses of the models of \citet{Woosley2017}, we find a good qualitative
agreement. His $36M_\odot$ model results in the ejection of $0.18M_\odot$ with a
kinetic energy of $3.7\times 10^{48}\;\rm erg$, and takes 18000 seconds from the
onset of instability until iron-core collapse. The $54M_\odot$ model of
\citet{Woosley2017} ejects $6.58M_\odot$ in four pulses, with a total kinetic energy of
$9.4\times10^{50}\rm\;erg$ and takes $150$ years from the onset of pulsations to
iron-core collapse. Except for the time to collapse of the more massive model,
all these numbers match within a factor of a few to our results, which is remarkable given how
steeply they change with the mass of the progenitor and the different initial
conditions used. For example, the $56M_\odot$ simulation of \citet{Woosley2017}
takes $\sim 1000$ years to undergo iron-core collapse from the beginning of the
pulsational phase.

\subsection{Grid of models}

We compute models of non-rotating pure helium stars in the range of initial masses
$M_{\rm initial}=40-100M_\odot$ at intervals of $2M_\odot$, with a finer mass
resolution near the edges to better resolve the minimum mass for a PPISN to occur and the
boundary between PPISN and PISN. For completeness, we also include models with
initial masses $>100M_\odot$ to resolve the upper mass limit at which BHs are
formed again.
These are summarized in
table \ref{tab:models}. The lower mass model of
$40M_\odot$ undergoes regular iron-core collapse and no eruptions, while the
$89.05M_\odot$ model is completely disrupted in a PISN.
PPISNe models with initial masses between $45M_\odot$ and $48M_\odot$
experience small amplitude pulsations and never restore their hydrostatic equilibrium,
as exemplified by the $46M_\odot$ model shown in Figure \ref{fig:pulses46}. All
other PPISNe models manage to restore hydrostatic equilibrium after the first
pulsation and have phases of quiescence before iron-core collapse. All PPISNe
and PISN models undergo the first pulse with $\sim 0.5M_\odot$ of helium left. Despite strong
mass loss, during core helium burning the convective core recedes
as mass is lost, and the time between core-helium depletion and the onset of the
SN is never enough to fully remove the remaining helium in the outer layers.

For models at zero
metallicity and without mass loss, \citet{Woosley2017} finds a range for
occurrence of PPISNe between $34-62M_\odot$. This can be compared to the range of
pre-SN masses, $M_{\rm pre\;SN}=35.1-60.8M_\odot$, at which we find
PPISNe. Despite our models being at a finite metallicity of $Z_\odot/10$ and
including mass loss, we see that both the lower and upper mass limits for the
occurrence of PPISN we obtain are consistent with those of \citet{Woosley2017}.

\begin{figure}
   \includegraphics[width=\columnwidth]{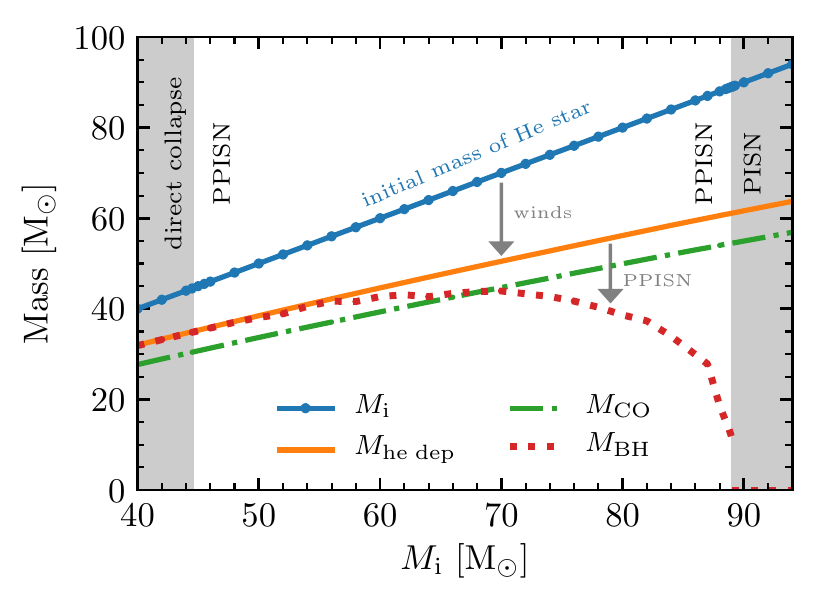}
   \caption{Masses at different evolutionary stages for PPISN models as a function
   of the initial mass $M_{\rm i}$. $M_{\rm
   He\;dep}$ and $M_{\rm CO}$ are the total and CO-core masses at helium
   depletion, while $M_{\rm BH}$ is the final mass of the BH formed. Individual dots in the blue $M_i$ line
   indicate individual simulations that were performed.} \label{fig:endmass}
\end{figure}

We show the resulting BH masses for our PPISN simulations in Figure \ref{fig:endmass}.
Models up to $M_{\rm i}=56M_\odot$ ($M_{\rm pre SN}\simeq 42
M_\odot$) undergo pulsations for less than a week and remove less than $1\%$
of the mass of the star prior to iron-core collapse,
resulting in only a small change in the final BH mass.
Models above $M_{\rm i}=70M_\odot$ ($M_{\rm pre SN}\simeq
50M_\odot$) lose more than $10\%$ of their mass through pulsations and take
between hundreds to ten thousand years between their first pulse and iron-core
collapse. These stars eject a significant fraction of their CO cores, resulting in a
monotonically decreasing $M_{\rm BH}$ as a function of $M_{\rm i}$.

We find the boundary between PPISN and PISN to be between our models with pre-SN masses
of $M_{\rm pre\;SN}=60.79M_\odot$ and $60.81M_\odot$ ($M_{\rm i}=89M_\odot$ and
$89.05M_\odot$), with the $M_{\rm pre\;SN}=60.79M_\odot$ star resulting in a
$\sim12M_\odot$ BH. We find that the final BH mass cannot be made arbitrarily
small by considering models closer to the PISN limit;
the inner $\sim10M_\odot$ of the $M_{\rm pre
; SN}=60.81M_\odot$ simulation actually reaches
hydrostatic equilibrium after the pulse, but is finally disrupted on a longer
timescale by the decay of radioactive nickel produced during the pulse.
What is important to emphasize here is that there is a physical
process that sets a non-zero value for the minimum remnant mass a PPISN can produce.
The particular values of $M_{\rm pre\;SN}=60.79M_\odot$ and $60.81M_\odot$
simply illustrate the sharp transition between PPISN and PISN, but should not be
interpreted as resolving the threshold to within $0.02M_\odot$.
As we show in Appendix \ref{app:conv}, our choice of a 21-isotope
nuclear reaction network can produce $\sim 10\%$ errors in nickel yields when
compared to more detailed networks, and the exact value for the transition
between PPISN and PISN will be modified by this.

\begin{figure}
   \includegraphics[width=\columnwidth]{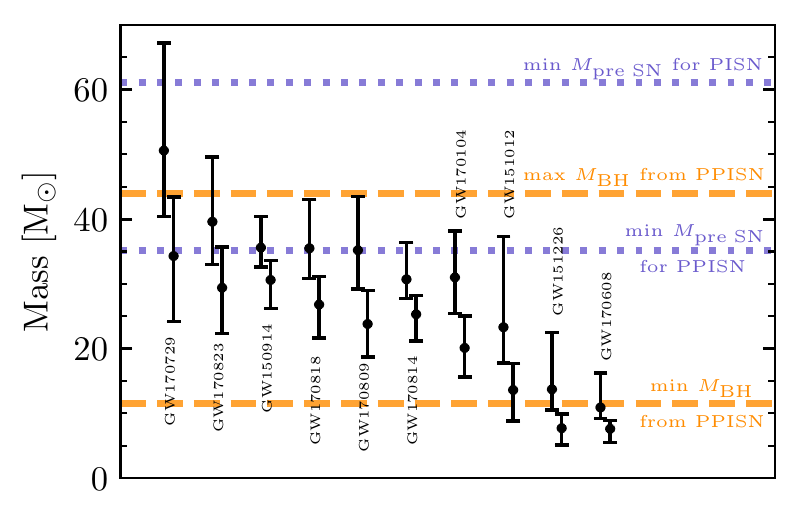}
   \caption{
   individual BHs in observed BBH mergers from the first Gravitational Wave
   Transient Catalog, ordered by total mass. Dotted horizontal lines
   indicate the range in $M_{\rm pre\;SN}$ for the occurrence of a PPISN in our
   models, with stars just above this range resulting in total disruption in a
   PISN. Dashed lines indicate the range of BH masses
   produced by PPISN. \label{fig:LIGO_bh_masses}}
\end{figure}

Figure \ref{fig:LIGO_bh_masses} shows the masses of individual BHs
for all BBH mergers observed so far, as reported in the first
Gravitational Wave Transient Catalog \citep{GWTC1}.
In particular, within the 90\% confidence intervals the more massive BH in GW170729 has a mass that exceeds the
minimum $M_{\rm pre\; SN}$ required for a PPISN, and even reaches beyond the
minimum $M_{\rm pre\; SN}$ required for a PISN. Within the large uncertainties
reported in individual masses, all
primary BHs except those of GW151226 and GW170608 have masses larger than the
minimum $M_{\rm pre\;SN}$ required for a PPISN, making them consistent from being the
product of a weak PPISN event. Since for higher mass PPISN progenitors the
final BH mass can be as low as $\sim12 M_\odot$, all of the BHs measured in BBH
mergers (except for the secondaries in GW151226 and GW170608) could be the
product of the evolution of $M_{\rm initial}>70M_\odot$ helium cores with high PPISN
mass loss. However, we expect this to be unlikely, as the initial mass function
disfavors such massive progenitors and they are more likely to be the result of
collapse from lower mass progenitors. We obtain an upper limit of $\sim
44M_\odot$ for the mass of BHs formed through PPISNe, which is in good agreement
with the results from the LIGO-Virgo collaboration that find that the observed
sample is well described by models that have $<1\%$ of BHs with masses above
$45M_\odot$ \citep{LIGOpop}.

\begin{figure}
   \includegraphics[width=\columnwidth]{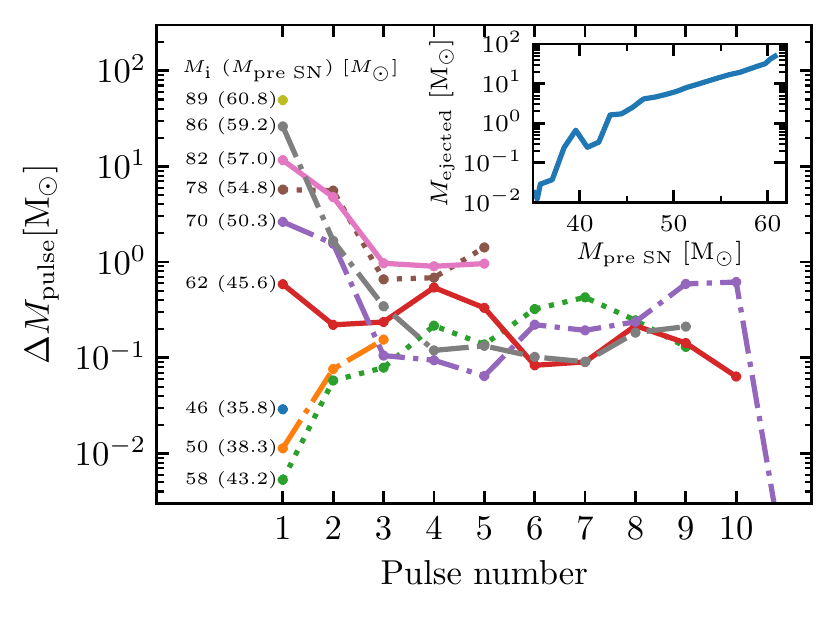}
\caption{Mass lost in individual pulses for a few representative
   models.\label{fig:pulsemass}}
\end{figure}
\begin{figure*}[ht!]
   \begin{center}
   \includegraphics[width=0.8\textwidth]{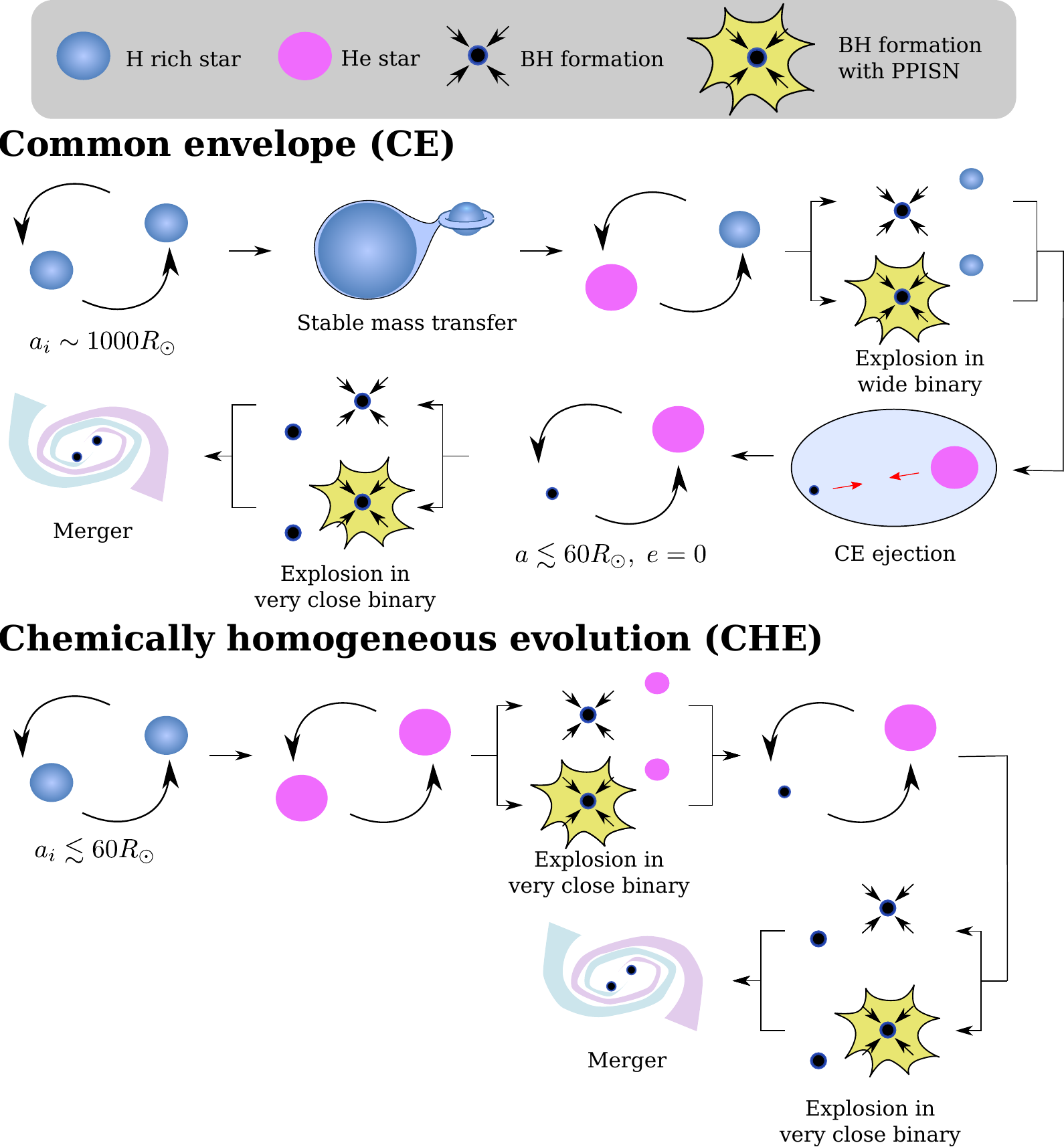}
   \end{center}
\caption{Possible occurrence of PPISNe in two different formation channels for
   merging BBHs through field binary evolution. The limit $a\lesssim 60$
   is required for a $40M_\odot+40M_\odot$ BBH to merge in less than $13.8$ Gyr,
   while separations $a>1000R_\odot$ are typical in the formation scenario of
   BHs similar to GW150914 through common envelope evolution (cf. \citealt{Belczynski+2016}).\label{fig:channels}}
\end{figure*}

We dissect the individual mass loss from each pulse in Figure
\ref{fig:pulsemass}.
Lower mass stars experience progressively larger pulses,
while the opposite is the case for the more massive systems. In addition to
this, the more massive models experience a long period of quiescence (up to tens
of thousands of years) between the
first and the second pulse (see Table \ref{tab:models}). It is these long-lived
objects that we focus on in the following section.

\section{Impact of close companions in a PPISN}\label{sec:ppisn2}
If merging BBHs can be formed by binaries in the field, we expect PPISNe
from hydrogen depleted stars to occur at different stages in their evolution if
they involve massive enough stars. This is illustrated in figure \ref{fig:channels}
for two different cases: (i) CE evolution in wide binaries
and (ii) CHE in very close binaries.
In both cases a BBH can be formed where either one or both components underwent
a PPISN. For the CE channel, a PPISN can happen with a companion in a wide orbit ($a\sim 1000R_\odot$) if
it collapses before the envelope is ejected through a CE, or in a compact orbit
($a\lesssim 60R_\odot$) if it happens after envelope ejection. In the case of
CHE, two PPISNe
from hydrogen free progenitors in a compact orbit are possible.

So far we have considered PPISNe to be unaffected by a nearby binary
companion. However, during a pulse heat is deposited throughout the entire star,
causing the post-pulse remaining layers to have a much more extended radius than
the starting object. We focus here on the systems that have long lifetimes after
their first pulse, studying the evolution of models with $M_i\geq 70M_\odot$ ($M_{\rm
pre\;SN}\geq 50.3$) which are quiescent for more than a century between the
first and the second mass ejection.

\subsection{Interaction right after a pulse}\label{sec:CE}

We first consider interaction happening
immediately after a pulse in a very close binary, close enough for the resulting
system to merge from the emission of GWs in less than the age of
the universe, $13.8$ Gyr \citep{PlanckColl2016}. For this
purpose, we take as a characteristic companion a $40M_\odot$ star or BH
(characteristic of the BH masses resulting from PPISN, see Table \ref{tab:models}) at a
separation of $a=58.6R_\odot$. This corresponds to the minimal
separation required for a $40M_\odot+40M_\odot$ BBH to merge in less than $t_m=13.8\;\rm
Gyr$ \citep{Peters1964}. Even accounting for enhanced eccentricity due to the
mass ejection, the final distance at periastron has to be $\le 58.6R_\odot$ for
a merger to happen within $13.8\;\rm Gyr$, so it can be used as an upper limit
to determine if a binary close enough to merge from GW
emission would interact after the pulse.

\begin{figure}
   \includegraphics[width=\columnwidth]{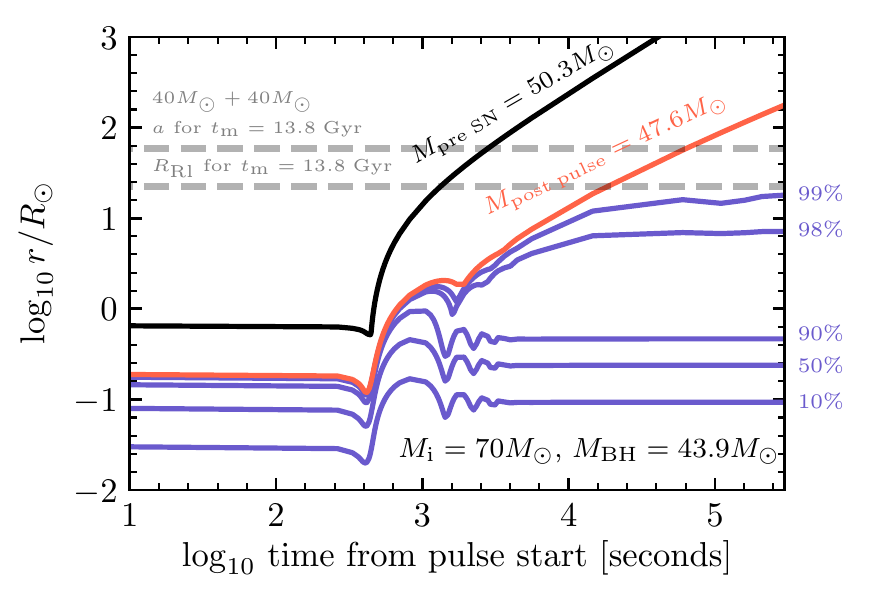}
   \includegraphics[width=\columnwidth]{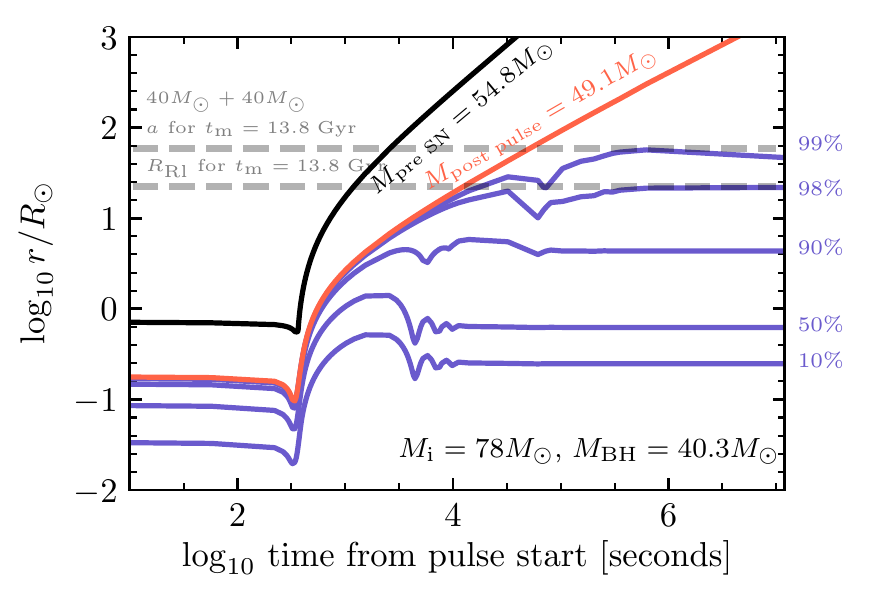}
   \includegraphics[width=\columnwidth]{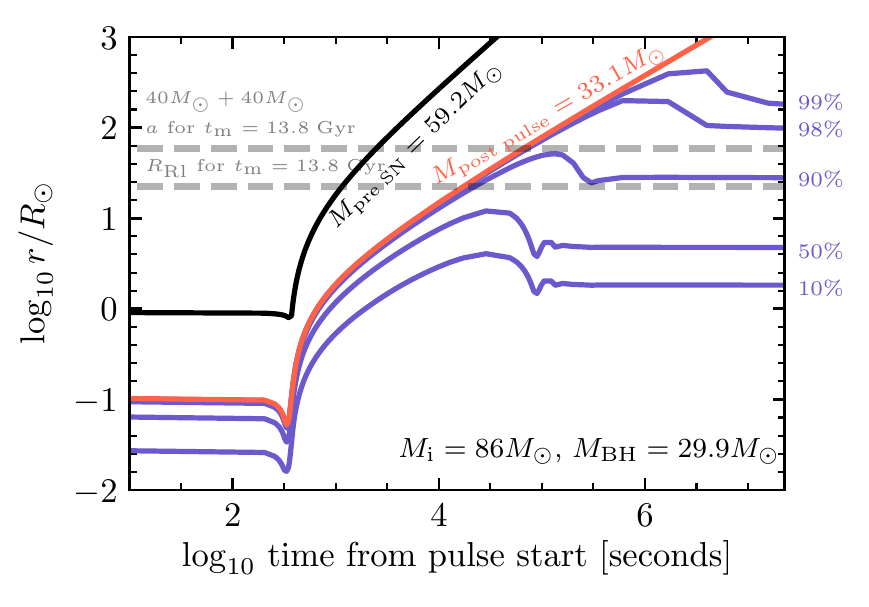}
   \caption{Evolution of different mass coordinates of two stars through the first
   PPISN pulse. Black lines indicate the pre-SN mass of the pulsating star,
   while orange lines indicate the remaining mass after the pulse (the mass
   coordinate where $v<v_{\rm esc}$). Purple lines indicate fixed mass fractions
   of the mass remaining after the pulse. For reference, the orbital separation
   ($a$) and Roche lobe radius ($R_{\rm Rl}$)
   for a $40M_\odot+40M_\odot$ BBH to merge due to GW radiation
   in $13.8\;\rm Gyr$ are shown with horizontal dashed gray lines. At the end
   points shown for each of these simulations, $99\%$ of the mass that remains
   bound is in hydrostatic equilibrium, and the outer layers are removed as
   described in Appendix \ref{app:relax}.\label{fig:pulse_evol}}
\end{figure}

The radial evolution through the first pulse of three of our simulations is
shown in Figure \ref{fig:pulse_evol}. An $M_i=70M_\odot$ progenitor
has a radius below $1R_\odot$ before the first pulse.
After the pulse, the outer layers expand significantly, in particular
the radius at a mass coordinate corresponding to $99\%$ of the mass that remains
bound expands by two orders of magnitude. At the end of the phase shown in Figure
\ref{fig:pulse_evol} there are $0.12M_\odot$ that extend beyond $r=58.6R_\odot$,
such that the remaining star would start interacting with a binary companion
close enough to result in a BBH merger.
The $78M_\odot$ and $86M_\odot$ models present even more extreme behaviour, with
the pulse resulting in $0.41M_\odot$ and $1.3M_\odot$ remaining beyond our nominal
choice of $a=58.6R_\odot$ at the end of the pulse. If these systems are to result in a merging BBH, then
they should exhibit strong interaction, possibly evolving into a CE immediately after the
pulsation.

Even if a significant amount of bound mass extends to regions beyond the orbital
separation, it is not obvious that the resulting system will
undergo an inspiral inside a CE. In particular, the time available
before iron-core collapse could be larger than the timescale for an inspiral due
to frictional drag. To see if this is the case, we consider the models at the
end-points of Figure \ref{fig:pulse_evol} and follow \citet{Taam+1978}
to estimate the energy dissipation rate due to the drag as
\begin{eqnarray}
   L_{\rm drag} = \pi R_A^2\rho v_{\rm rel}^3,
\end{eqnarray}
where $R_A$ is the accretion radius, the density $\rho$ is taken at the radial coordinate $r=58.6R_\odot$ of
the post-pulse star and $v_{\rm rel}$ is the relative velocity of the
inspiraling companion and its surrounding envelope. For simplicity we consider a circular orbit with a
separation $a=58.6R_\odot$ and component masses $M_1=M_{\rm post\;pulse}$ and
$M_2=40M_\odot$. Assuming the rotation velocity of the expanded layers is
negligible since they rapidly expand by about two orders of magnitude, the
relative velocity is simply the sum of the orbital velocities of both
components, and is of the order of $\sim 500\;\rm km\;s^{-1}$ for the three
models we consider. The accretion radius can be computed as
\begin{eqnarray}
   R_A=\frac{G M_2}{v_{\rm rel}^2+c^2},
\end{eqnarray}
where $c$ is the local sound speed. We find $R_A$ to be on the order of $\sim 20R_\odot$. Since this is comparable
to the orbital separation, in order to provide a conservative estimate of the
drag we use instead $R_A=H_P$, the local pressure scale height of the star at
$r=58.6R_\odot$, which we find to be $\sim 10R_\odot$ for these three models. The
characteristic timescale for inspiral can then be estimated as
\begin{eqnarray}
   \tau_{\rm ins}=\frac{a}{|da/dt|},\qquad \frac{da}{dt}=\frac{L_{\rm
   drag}}{GM_1M_2/2a^2}.\label{equ:tauins}
\end{eqnarray}

\begin{figure}
   \includegraphics[width=\columnwidth]{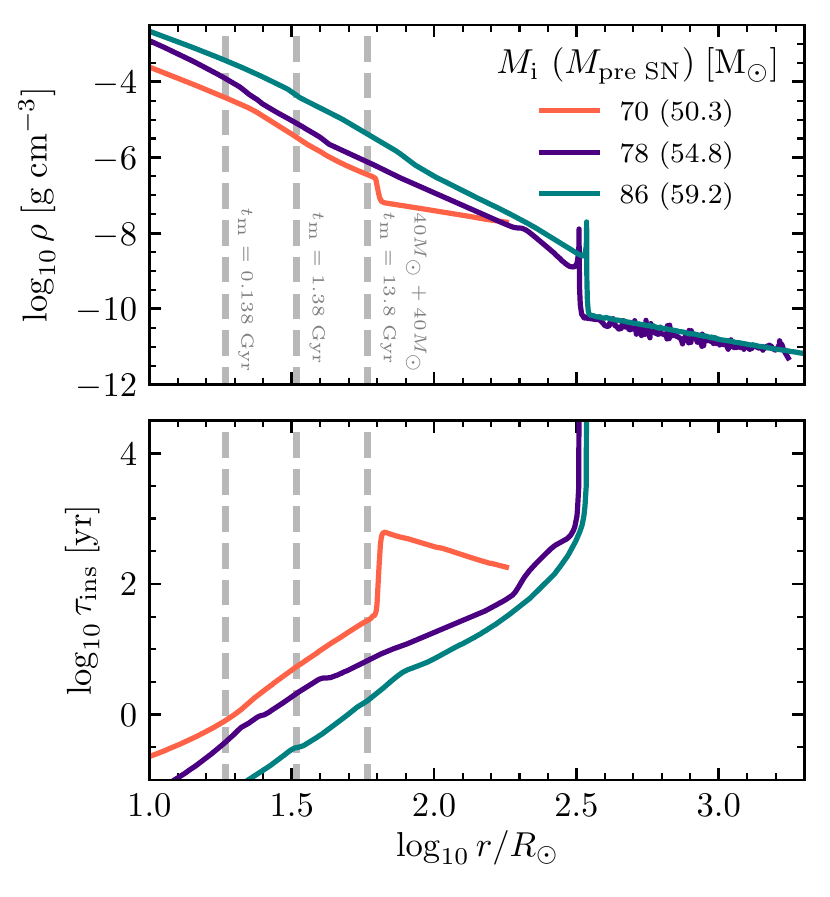}
   \caption{(top) Density profiles of layers with velocities $v<v_{\rm esc}$
   after the end of the first pulse for the three PPISN models shown in Figure
   \ref{fig:pulse_evol}. Dashed vertical lines indicate the separation required
   for a $40M_\odot+40M_\odot$ BBH to merge due to emission of GWs in $13.8$,
   $1.38$ and $0.138$ Gyrs. (bottom) Inspiral timescale assuming a $40M_\odot$
   companion at a circular orbit with separation $a=r$ (see Eq. (\ref{equ:tauins})).
   \label{fig:inspiral_tau}}
\end{figure}

Computing this for our $70M_\odot$, $78M_\odot$ and $86M_\odot$ progenitors
results in $\tau_{\rm ins}=28,\;6.7$ and $1.6$ years respectively.
Since these stars are expected to live for more
than a century before additional pulses and iron core-collapse occurs, there is enough time for an inspiral to
happen. Figure \ref{fig:inspiral_tau} shows how these results are modified by a different
choice of orbital separation. In particular, for the $78M_\odot$ and $86M_\odot$
models, which have a
lifetime $>1000$ yrs after the first pulse, successful inspirals are expected
even up to radii an order of magnitude larger than $a=58.6R_\odot$. Thus, the
development of a CE inspiral is expected to happen for a wide range
of separations.

Estimating the outcome of these inspirals is much more uncertain using our 1-D
models, considering the star at this point has ejected almost all its helium and
it is an extended CO core with no well defined core-envelope boundary. This adds
to all the uncertainties associated to CE
evolution (cf. \citealt{Ivanova+2013}).
Despite the short inspiral timescales, the orbital separation and eccentricity
will not necessarily be significantly affected by the common envelope.
This is because a small reduction in the
separation can provide enough energy to remove the
relatively small amount of mass in the extended layers.

\begin{figure}
   \includegraphics[width=\columnwidth]{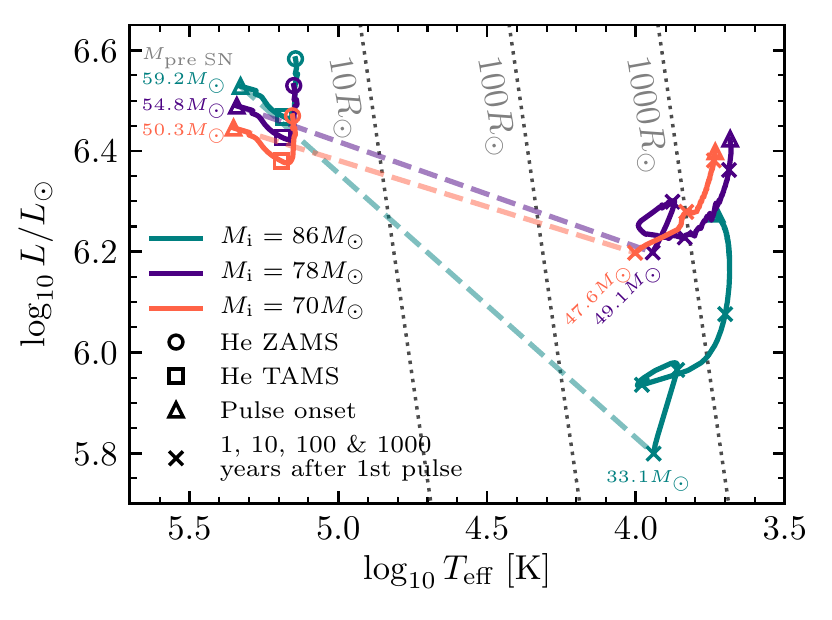}
   \caption{HR diagram showing the evolution of helium star models with $M_i=70M_\odot$,
   $78M_\odot$ and $86M_\odot$. Dashed lines connect the properties of the star
   at the onset of the first pulse, and one year afterwards, with
   crosses indicating periods of time 1, 10, 100 and a 1000 years after the
   onset of the first pulse. Evolution after the onset of the second pulse is not shown, but
   corresponds to less than $1$ yr in time before iron-core collapse. All models
   evolve to become CO red supergiants.\label{fig:hr_diagram}}
\end{figure}

\pagebreak

\subsection{Interaction during inter-pulse phases}
Even if the PPISN happens in a wide binary with $a>1000R_\odot$, we expect
interaction to happen. Figure \ref{fig:hr_diagram} shows tracks in the
HR diagram of the same $M_i=70M_\odot$, $78M_\odot$ and $86M_\odot$ progenitors
we discussed above, including the evolution before the first pulse, and between
the first pulse and the second. During the evolution after the first pulse
the ejected layers have been removed following the procedure described
in Appendix \ref{app:relax}, so the luminosity and effective temperature shown
correspond to the photosphere of the bound star that is left.

As convection develops in the outermost layers
of these stars, they expand to become red supergiants with radii in excess of
$1000R_\odot$. These objects are quite peculiar, as through the pulsation all
the helium rich layers are ejected, resulting in a red supergiant
composed almost entirely of carbon and oxygen at its surface.
This expansion will result in Roche lobe overflow even for
binaries at a separation $\sim 3000R_\odot$.
If this happens, mass transfer could be either stable or unstable depending on
the mass ratio of the system and the response of the donor star
to mass loss \citep{Soberman+1997}. Unstable mass transfer would proceed on a
dynamical timescale, and could lead to a CE inspiral.
If mass transfer is stable, we expect it to operate on the same timescale that the star is expanding.
This is the thermal timescale of this extended envelope, which is of
the order of $\sim 10000$ years, such that subsequent PPISN pulses would happen
while the star is still transferring mass to a companion. In the context of the
CE formation channel of merging BBHs, the companion at this point would be a
non-degenerate star most likely on the main sequence. 

\section{Impact of PPISN on merging BBHs}\label{sec:mergingBBH}

\subsection{Change in spin}\label{sec:spinchange}
Observationally, the spin of a merging BBH is constrained mostly in terms
of the parameter $\chi_{\rm eff}=(m_1\chi_1+m_2\chi_2)/(m_1+m_2)$, where $m_1$ and
$m_2$ are the individual masses of each BH and $\chi_1$ and $\chi_2$ their
projected spin parameters on the orbital plane. There is an important
degeneracy between $\chi_{\rm eff}$ and the mass ratio of the merging BBH,
which limits the precision to which each can be measured independently
\citep{Hannam+2013}. Despite this the measurements so far by the aLIGO and aVirgo
detectors have shown that $\chi_{\rm eff}$ is centered around zero
\citep{Abbott_LIGOsummary_2016}, indicating that either the BH spins are small,
or significantly misaligned with the orbital plane. 

One potential source of orbit misalignment are kicks from
asymmetric pulses and their associated mass loss. To our knowledge, there are no multi-D simulations assessing how
symmetric PPISNe ejections are. \citet{Chen+2014} performed 2-D simulations of
colliding shells from a PPISN but did not model their actual ejection,
so it does not provide information on potential kicks produced on the remnant.
For this discussion we ignore the effect of kicks on $\chi_{\rm eff}$, but note
that they would only result in a reduction of it.

The spin parameter of a BH will depend on the distribution of angular momentum
in its progenitor. Stellar winds are an efficient mechanism to remove angular momentum from a star
(cf. \citealt{Heger+2005}), as the long timescales involved allow for
efficient coupling between the stellar envelope and its core.
PPISN eruptions can
remove a large fraction of the mass of a star, but in contrast to wind mass loss
they happen in a dynamical timescale of the star preventing efficient coupling.

As we only consider non-rotating stellar models, we cannot self-consistently measure the
impact of eruptions on the final BH spin. However, it can be approximated under
a few assumptions. Consider the spin parameter at mass coordinate $m$,
$a(m)=J(m)c/m^2G$, where $J(m)$ is the angular momentum contained inside the
mass coordinate $m$.
If the star rotates as a solid body with an angular frequency $\omega$,
then $a(m)\propto \omega I(m)/m^2$, where $I(m)$ is the moment of inertia of the
star up to $m$. A rapid mass loss event that reduces the mass of the star from
$M_{\rm pre\;SN}$ to $M_f$ then produces a relative change in the spin of
\begin{eqnarray}
   \frac{a(M_{f})}{a(M_{\rm pre\;SN})}=\frac{I(M_{f})}{I(M_{\rm
   pre-SN})}\frac{M_{\rm pre\;SN}^2}{M_{f}^2}. \label{equ:spindown}
\end{eqnarray}
If the amount of mass loss during a PPISN does not depend
strongly on rotation at the moment of collapse, we can use the final BH mass
$M_{\rm BH}$ predicted by our models to compute the relative change in spin.

It has to be pointed out that there are clear caveats to this calculation, in particular for the
case of PISN it is known that their evolution can be altered by rapid rotation,
as progenitors can be stabilized due to centrifugal forces and lead to weaker
explosions \citep{Glatzel+1985,Chatzopoulos+2013}. Also, during late burning stages,
the inner regions of a star are expected to decouple
and rotate at higher angular frequencies than the outer layers (cf.
\citealt{Heger+2000}). These two effects are expected to reduce the angular
momentum lost through eruptive mass loss, since they imply less mass loss and
that the assumption of solid body rotation overestimates the angular momentum of
the outer layers. The estimate given by Equation (\ref{equ:spindown}) then represents
the maximum effect PPISN mass loss can cause to the final BH spin.

\begin{figure}
   \includegraphics[width=\columnwidth]{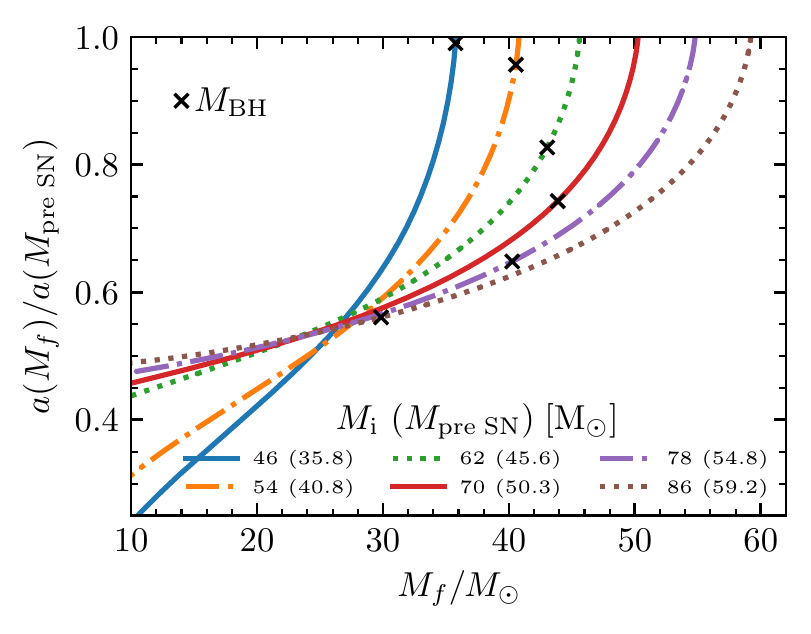}
   \caption{Final spin in terms of its initial one for eruptive mass loss. Each
   line indicates one of our models on the onset of collapse, and the black
   crosses indicate the final mass of the BH formed after undergoing a PPISN.\label{fig:am_change}}
\end{figure}

Figure \ref{fig:am_change} shows the result of computing Equation
(\ref{equ:spindown}) for some representative models in our grid spanning the
entire PPISN range. Most systems only experience
reductions of $\sim 30\%$, and it can be seen that even for stronger pulses the spin
cannot be reduced below 50\% of its initial value. Although they produce a
non-negligible change in the spin, PPISN eruptions are not be capable of
reducing the effective spin of a progenitor with $\chi_{\rm eff}\sim 1$ down to the values observed by
aLIGO/aVirgo. Kicks produced during PPISNe could further reduce
$\chi_{\rm eff}$ by misaligning the orbit and the spin of the BH, but whether or
not PPISNe could lead to strong kicks is uncertain.

\subsection{Eccentricity enhancement}
It is expected that
the upcoming LISA observatory will detect GWs from inspiraling
BBHs up to years before they are detectable by ground-based observatories
\citep{Sesana2016}. This opens up the possibility of measuring eccentricities
for these sources, which can be used to distinguish between formation scenarios
\citep{Nishizawa+2016,Breivik+2016}. In particular, dynamical formation
scenarios can produce highly eccentric BBHs \citep{Rodriguez+2016a,
Antonini+2017}, allowing
them to be distinguished from BBHs produced through field binary evolution.
However, dynamical ejections of mass in field binaries can also change the
eccentricity of these systems \citep{Blaauw1961,Boersma1961}.

In order to estimate if mass loss through PPISN can produce systems with measurable
eccentricities in the LISA frequency band, we consider two
different scenarios:
\begin{enumerate}
   \item The system is formed through CE evolution. In this case,
      the eccentricity induced by the first PPISN is erased by a CE
      phase. Only the second formed BH contributes to the final
      eccentricity when it undergoes a PPISN (see Figure \ref{fig:channels}).
   \item The system is formed through CHE. In the
      absence of a CE phase, PPISN from both stars contribute to the final
      eccentricity.
\end{enumerate}

For both cases, in order to compute eccentricities at the moment of
BBH formation we assume the following: 
\begin{itemize}
   \item Each mass ejection is
completely symmetric and imparts no momentum kick on the layers that remain
bound. We also ignore binary interaction in between pulsations and assume the
      material is ejected at a velocity much larger than the orbital velocity.
The resulting ejection is analogous to a Blaauw kick
\citep{Blaauw1961}, and
therefore produces a change in orbital eccentricity that is independent of orbital
separation.
   \item A PPISN can undergo multiple pulses
before collapsing, each affecting the orbital parameters in a different way
      depending on the orbital phase at the moment of the ejection.
      We assume the orbital phase at each pulse 
      has a flat distribution as the periods of quiescence for
PPISN models experiencing significant mass loss are $> 1$ yr, much longer than
      the orbital periods required for a GW merger within a Hubble time. For
      example, a $40M_\odot+40M_\odot$ BBH in a circular orbit must have an
      orbital period $<6$ days to merge in less than 13.8 Gyrs.
   \item We assume that at the moment each component undergoes a PPISN and forms
      a BH they are hydrogen-stripped stars with equal pre-SN masses ($M_{\rm
      pre\;SN}$), leading
      to an equal mass BBH. This does not imply that the initial mass ratio of
      the system was unity or that both stars explode simultaneously, but rather
      that binary interaction happens to produce near equivalent progenitors. Choosing unequal mass
systems leads to higher final eccentricities, so our assumption sets a
lower limit on the resulting eccentricities.
\end{itemize}
Under these assumptions, the initial orbital separation does not play a role in
the final eccentricity, and for an individual system one obtains a distribution of
eccentricities rather than a unique value.

As in Section \ref{sec:spinchange}, we assume here that PPISNe do not
result in a momentum kick on the resulting remnant.
Nevertheless, we have performed simple tests with non-zero kick velocities and
found that in general they produce distributions with higher eccentricities.
However, for simplicity, we only discuss the case for Blaauw kicks here, so our
results serve as a lower limit.

\begin{figure}
   \includegraphics[width=\columnwidth]{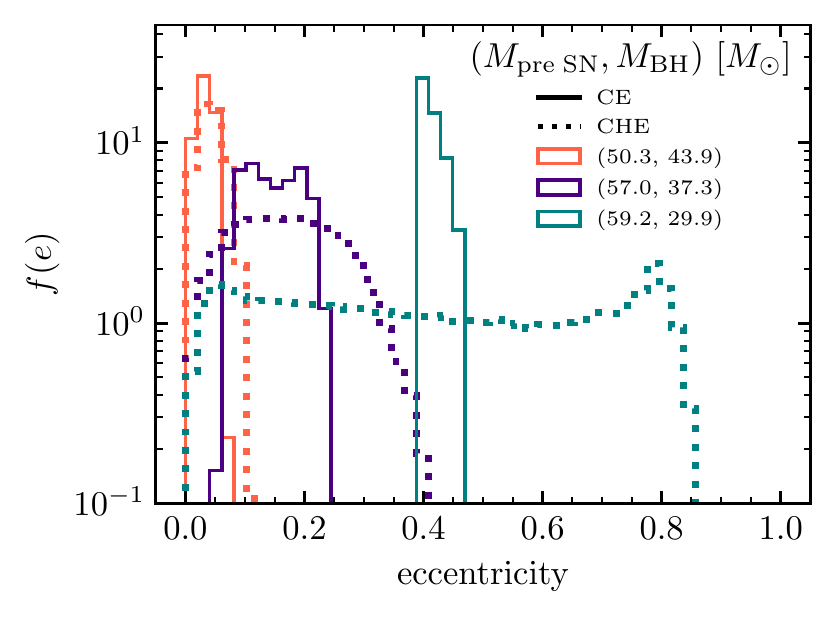}
   \caption{Example distributions of eccentricity enhancements produced by
   PPISN in binary systems from CE and CHE evolution. 
   Eccentricities shown correspond to the moment after the formation of
   the second black hole. We consider systems that
   would result in BBH mergers with a mass ratio of unity, with each color in
   the diagram indicating the pre-SN mass of the hydrogen depleted progenitor
   used and its resulting BH mass. \label{fig:eccentricity_dist}}
\end{figure}

Figure \ref{fig:eccentricity_dist} shows the resulting eccentricity distributions for
some of our higher mass models, computed using $10^5$ samples for each mass.
Lower mass PPISN progenitors do not lose enough mass to produce eccentricities
larger than $0.1$. More massive models can actually become unbound as they
eject more than half of the total mass in the system \citep{Blaauw1961}, but such extreme systems
only happen in a reduced mass range and we expect them to be uncommon.

\begin{figure}
   \includegraphics[width=\columnwidth]{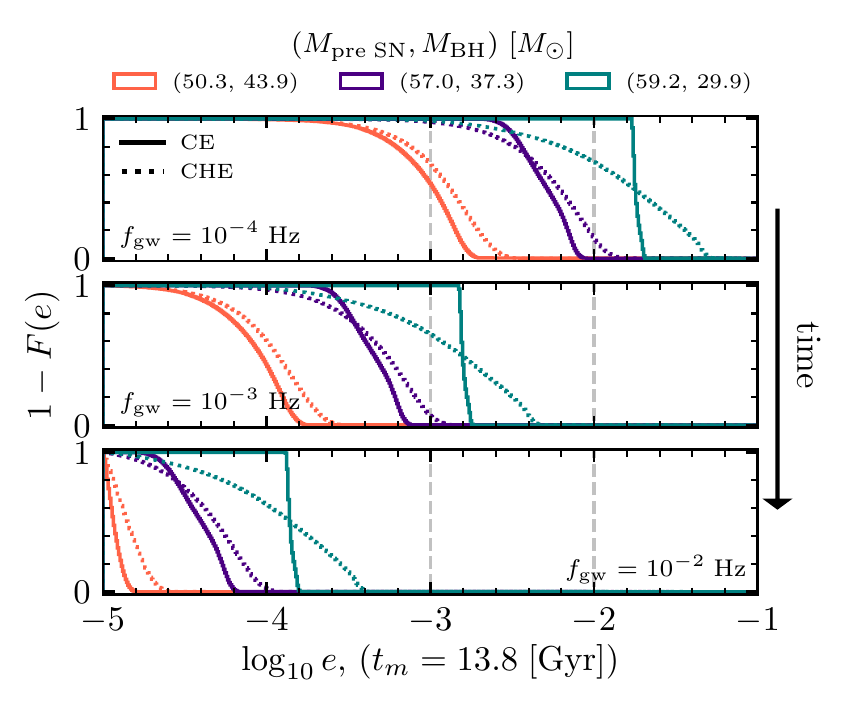}
   \includegraphics[width=\columnwidth]{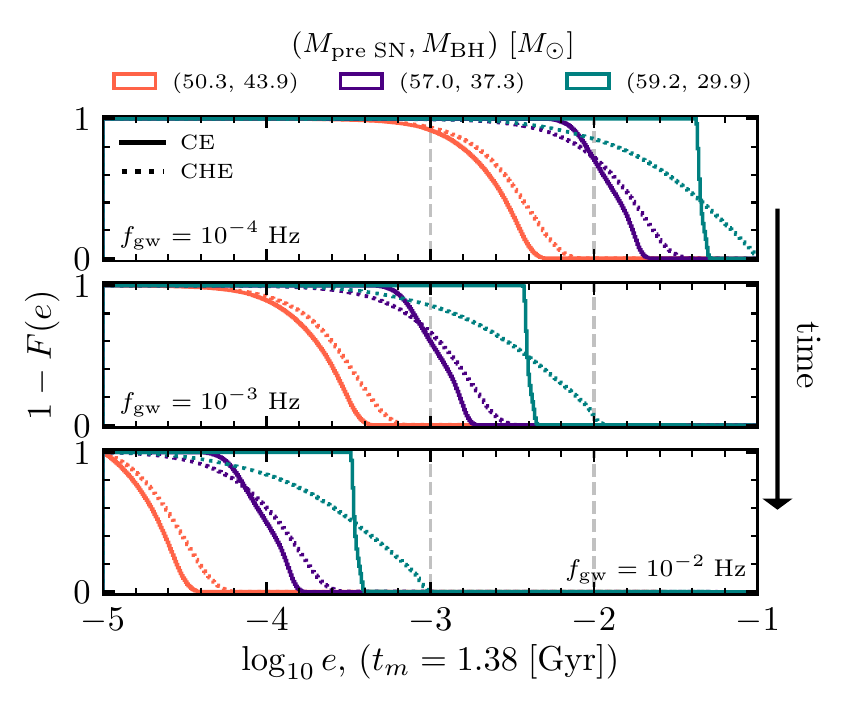}
   \caption{Inverse cumulative distribution functions for eccentricities of
   different BBH progenitors at
   frequencies for GW emission relevant to the LISA observatory.
   As time passes, a BBH is circularised due to the emission of GWs,
   and emits at higher frequencies.
   Top figure assumes that the merger time for all sources is $t_m=13.8$ Gyrs,
   while the bottom figure assumes $t_m=1.38$ Grys.
   \label{fig:LISAecc}}
\end{figure}

After the formation of a BBH, GWs will reduce the orbital period $P$ as well as the
eccentricity of the system. Both quantities then follow a relationship
$P=P(e)$ which is independent of the component masses \citep{Peters1964},
\begin{eqnarray}
   \begin{aligned}
      P(e)=P_0\left(\frac{A(e)}{A(e_0)}\right)^{3/2},\qquad\qquad\qquad\;\\
      A(e)=\frac{e^{12/19}}{(1-e^2)}\left(1+\frac{121}{304}e^2\right)^{870/2299},
   \end{aligned}
\end{eqnarray}
where $P_0$ and $e_0$ are the initial values. As the orbital period is reduced,
the frequency of GW radiation $f_{\rm GW}=2/P$ increases.
This means that to translate the birth
eccentricities we have computed into eccentricities in the LISA band, we need to
specify a birth period as well. As an extreme choice, we set the initial period
for each of our simulated binaries such that they have a merger time of $t_m=13.8$
Gyr, and also consider the case when $t_m=1.38$ Gyr instead. Figure
\ref{fig:LISAecc} shows how the distributions shown in Figure
\ref{fig:eccentricity_dist} are changed as a binary evolves due to GW radiation under these assumptions.

\citet{Nishizawa+2016} studied the expected accuracy for eccentricity
measurements with the LISA observatory, considering the case of detected BBHs that merge
within the mission lifetime $T_{\rm obs}$. For the case of a merger of two
$40M_\odot$ BHs, this requires the source to be emitting at a frequency larger
than $10^{-2}$ Hz at the beginning of the mission. They showed that
eccentricities in excess of $e>0.01$ would always be measured by
LISA, while eccentricities $e>0.001$ can be measured for
$90\%(25\%)$ of those mergers considering  $T_{\rm obs}=$5 yrs (2 yrs). None of our PPISN
models reach the peak of sensitivity of LISA ($\sim 10^{-2}$ Hz) with eccentricities above
0.001 so we expect them to be below the threshold for detectability. Thus, we still expect this
population to be distinguishable from BBHs predicted to form through dynamical
formation.

Note however that there is a big caveat to these calculations. As we have shown
in Section \ref{sec:ppisn2} the systems that we expect to produce
measurable eccentricities by LISA are the same ones that would interact strongly
in the centuries to millennia long phases between pulsations. The calculations
done here assume that no circularisation due to either tidal interactions or CE
evolution happens during this period, something
that requires further work to properly assess. As discussed in
Section \ref{sec:CE} for CE phases occurring immediately after the first PPISN
pulse, despite the inspiral timescales being short compared to the remaining
lifetime of the pulsating star this does not imply that the orbital separation
and eccentricity will be significantly affected by the CE event.

\subsection{Impact on chirp masses}\label{sub:pop}
To study how PPISN would affect measured chirp masses ($M_{\rm chirp}=[m_1
m_2]^{3/5}/[m_1+m_2]^{1/5}$) of merging BBHs, we
develop a simple model which does not assume any particular
formation scenario. \citet{Abbott_LIGOsummary_2016} assume that the more
massive BH from a merging BBH follows a Salpeter law $dN/dM_{\rm
BH,1}\propto M_{\rm BH,1}^{-\alpha}$, and that the masses of secondaries follow a
flat distribution ranging from $M_{\rm min}=5M_\odot$ to $M_{\rm BH,1}$. In a
similar way, we assume the pre-SN mass of one star, $M_{\rm pre\;SN,1}$,
follows a Salpeter distribution, and that its companion mass is distributed flat
between $\max(5M_\odot, 0.5M_{\rm pre\;SN,1})$ and $M_{\rm pre\;SN,1}$. This
limits the mass ratio before BH formation to be above $0.5$, and is motivated by
most formation channels clearly favoring mass ratios closer to unity (cf.
\citealt{Dominik+2012}, \citealt{Rodriguez+2016a}, \citealt{Marchant+2016},
\citealt{Chatterjee+2017}).

We randomly sample these distributions, and for each
star, if its mass falls below the range for PPISN of our grid, we assume it
collapses directly to form a BH of mass $M_{\rm pre\;SN}$. On the contrary, if it
falls above the range of our PPISN models, we assume it is completely disrupted in a
PISN. In the range in-between, we interpolate our grid to obtain the final mass
of the remnant BH. For $\alpha$, we choose $2.35$ which corresponds to a
Salpeter initial mass function \citep{Salpeter1955}. This value of $\alpha$ is
consistent with the the value inferred by
\citet{Abbott_LIGOsummary_2016} using the observed BBHs in the first aLIGO
observing run and assuming a power law distribution. The objective of this experiment is not to provide a definitive
prediction, but just to illustrate how much of an effect PPISN can have under
simple assumptions on the progenitor population. Note that in this simple approach
we do not consider the increase in merger time that would result from ejections.
It is not clear if this would bias observations for or against systems which
underwent PPISNe, as longer delay times can lead to mergers at smaller
redshifts.

\begin{figure}
   \includegraphics[width=\columnwidth]{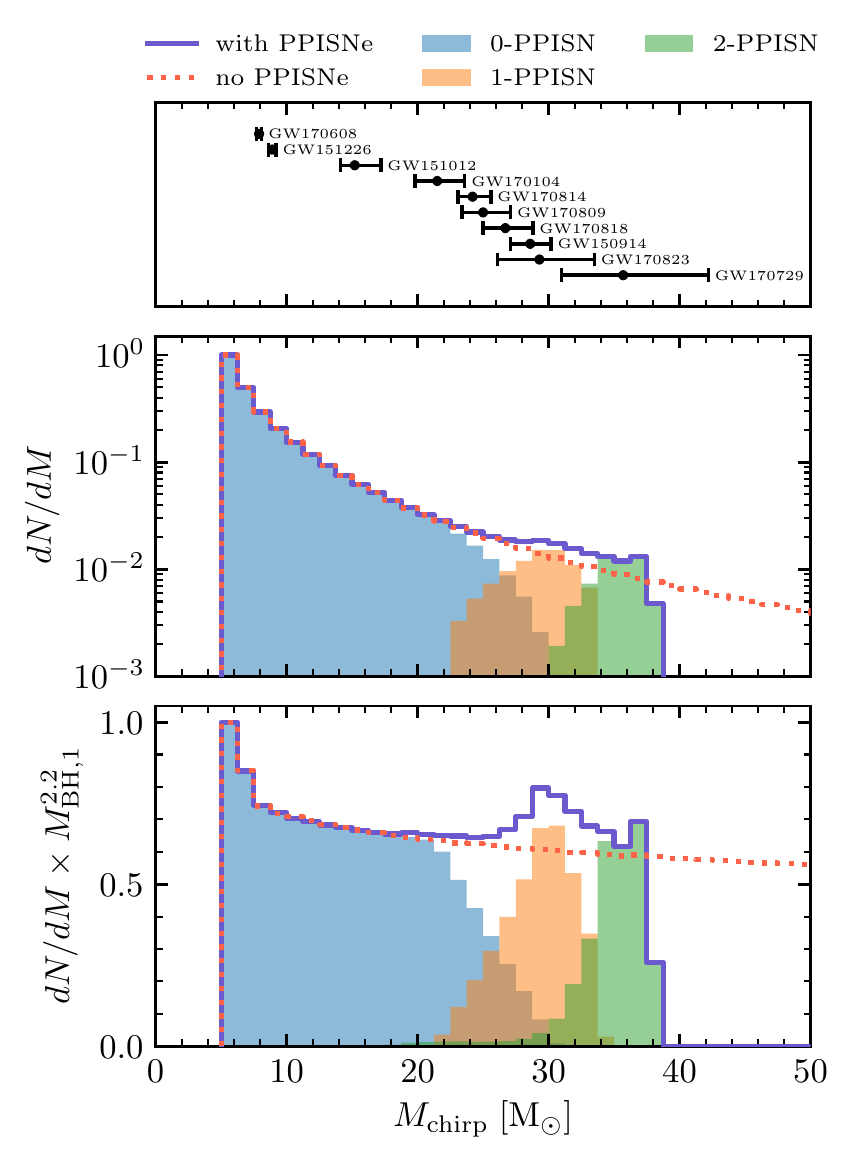}
   \caption{(top) Chirp masses of all observed BBH mergers from the
   first Gravitational Wave Transient Catalog. (middle) Distribution of chirp masses in our population synthesis calculation
   compared to a model with no PPISN. The distribution for models with PPISN is
   further separated on the contribution of systems which undergo two, one,
   or no PPISN before formation. (bottom) Same as before, but scaled by $M_{\rm
   BH,1}^{2.2}$ to roughly account for the sensitivity dependence with primary
   mass of the aLIGO/aVirgo detectors.\label{fig:chirp_dist}}
\end{figure}

The results of this calculation are shown in Figure \ref{fig:chirp_dist}. For
reference, we include a set of calculations where all BHs are assumed to form
through direct collapse. Systems that undergo either one or two PPISN
events result in lower chirp masses than the model without PPISN, producing a
pileup just below the PISN gap. Moreover, the sensitivity of the aLIGO detectors
scales roughly with $M_{\rm BH,1}^{2.2}$ up to total masses of $100M_\odot$
\citep{FishbachHolz2017}. Scaling the distribution of chirp masses we have
computed by this factor turns the distribution into a double peaked one. Future
observing runs of the aLIGO/aVirgo detectors are expected to observe tens of
merging BBHs in the coming years, constraining the distribution of their chirp
masses \citet{Abbott_LIGOsummary_2016,Abbott_2018_obsruns}. If a clear double-peaked structure comes out of these measurements, then
it should not necessarily be interpreted as two distinct formation channels.
Current observations do favor a dearth of BH masses in excess of
$45M_\odot$, but more observations are required to place constraints on the
existence of a peak produced by PPISN \citep{LIGOpop}.

\section{Conclusions} \label{sec:discussion}

We have shown that PPISNe can lead to strong binary interaction before iron-core
collapse and BH formation, with systems in orbits compact enough to result in
BBH mergers undergoing CE events after the first pulse. Although we do not know the outcome of such CE
phases, they can potentially provide interesting electromagnetic counterparts to
the PPISN itself. If there is a successful ejection of the CE, this is expected
to be observable as a luminous red novae \citep{Ivanova+2013b}.
Alternatively, the system could fail to eject the CE and result in a merger.
If the inspiraling object is a BH, given the large budget of orbital
angular momentum in the system conditions could be appropriate for a long gamma-ray burst in a
similar way to the collapsar scenario \citep{Woosley1993}. Even if the CE is
ejected, if a few solar masses of material fallback into a BH companion this can
provide sufficient energy to power a hydrogen-poor superluminous supernova
\citep{Moriya+2018}. Long-lived phases of Roche lobe overflow with a companion
BH would lead to the formation of ultraluminous X-ray sources (see
\citealt{Kaaret+2017} for a recent review), with peculiar CO giants or
supergiants as donors. Most of these potential outcomes are speculative at this
point, but merit detailed further study potentially through the use of 3D
hydrodynamical simulations.

We have also shown that PPISN can modify various observable properties of
merging BBHs, including their spins, eccentricities and chirp masses. However,
to do so we have ignored the potential interaction of a star
undergoing a PPISN with its companion. Properly characterising these interaction
phases is then fundamental to understand how stars that undergo PPISNe
contribute to the overall population of merging BBHs.

\acknowledgments
PM would like to thank Francis Timmes, Giacomo Terreran, Takashi Moriya and
Christopher Berry for useful discussions.
PM acknowledges support from NSF grant AST-1517753.
SdM and MR acknowledge funding by the European Union's Horizon 2020 research and innovation programme
from the European Research Council (ERC) (Grant agreement No.\ 715063), and by the Netherlands
Organisation for Scientific Research (NWO) as part of the Vidi research program BinWaves with project
number 639.042.728. RF is supported by an NWO top grant with project number 614.001.501.
VK is a CIFAR Research Fellow, and VK and KP acknowledge support by the CIFAR
program in Gravity and Extreme Universe.
PM also thanks the Kavli Institute for theoretical physics of the university of
California Santa Barbara, together with the participants of the "Astrophysics
from LIGO's First Black Holes" and "The Mysteries and Inner Workings of Massive
Stars" programs for helpful discussion.
For this project the SAO/NASA Astrophysics Data System (ADS) was used
extensively.

\software{
\texttt{MESA}
\citep[][\url{mesa.sourceforge.net}]{Paxton+2011,Paxton+2013,Paxton+2015,Paxton+2018},
\texttt{Python} avaliable from \href{https://www.python.org}{python.org},
\texttt{matplotlib} \citep{hunter_2007_aa},
\texttt{NumPy} \citep{der_walt_2011_aa},
\texttt{ipython/jupyter} \citep{perez2007ipython,kluyver2016jupyter},
\texttt{inkscape} available from \href{https://inkscape.org/}{inkscape.org}.
         }

\appendix 

\section{Time dependent convection} \label{appendix:tdc}
During phases of dynamical instability large regions in the star can switch back
and forth between being stable or unstable to convection on timescales
comparable to or shorter than a convective turnover timescale. To properly treat
energy transport under these conditions, a model for time-dependent convection
is required. Here we describe a simple model that captures the relevant
timescales and reduces to standard mixing-length theory (MLT,
\citealt{Bohm-Vitense1958}) in long
timescales. We follow the work of \cite{Arnett1969} and consider the average convective
velocity $v_c$ in MLT to be an independent variable which satisfies the equation
\begin{eqnarray}
   \frac{\partial v_c}{\partial t}=\frac{v_{\rm MLT}^2 -
   v_c^2}{\lambda},\quad\rm{for\;convectively\;unstable\;regions}
   \label{app:tdc_conv}
\end{eqnarray}
where $v_{\rm MLT}$ is the steady state value predicted by MLT. $\lambda$ is the mixing length,
which we define as $\alpha_{\rm MLT}H_P$ where $\alpha_{\rm MLT}$ is a free
parameter of order unity and $H_P$ is the local pressure scale height. In
particular for our simulations we use $\alpha_{\rm MLT}=2$. On
timescales much longer than a convective turnover timescale ($\tau_{\rm to}=\lambda/v_{\rm
MLT}$) convective velocities asymptotically approach the steady state value
$v_{\rm MLT}$, recovering standard MLT. In regions that are convectively stable
$v_{\rm MLT}=0$ and simply using Equation (\ref{app:tdc_conv}) would result in
convective velocities decaying on a timescale $\tau=\lambda/v_c$ which becomes
infinetely large as convective velocities are reduced. This ignores the
actual timescale in which fluid parcels would be slowed down in a stratified
medium. To provide an order of magnitude correction to this, we construct a
timescale $\tau_N=1/N$ where $N$ is the Brunt-V\"ais\"all\"a frequency and use
\begin{eqnarray}
   \frac{\partial v_c}{\partial t}= -\frac{v_c^2}{\lambda} - \frac{v_c}{\tau_N},\quad\rm{for\;convectively\;stable\;regions}
\end{eqnarray}
to model the shutoff of convection. Mixing from convection is
modeled as a diffusive process with a diffusion coefficient $D=v_c\lambda/3$.

In its standard form, MLT solves an algebraic system of three
equations to compute the steady state convective velocity $v_{\rm MLT}$, the
temperature gradient of the star $\nabla$, and the temperature gradient of
displaced blobs of material $\nabla'$, which differs from the adiabatic gradient
$\nabla_a$ due to radiative energy losses. In our case, we require a derivation
of MLT for a given value of $v_c$ rather than the steady state one.
Following \citet{CoxGiuli1968}, if convective velocities are given then the
convective efficiency $\Gamma$ (which is the ratio of energy radiated by a
moving parcel, to the energy released when it dissolves after crossing a mixing
length) can be directly computed as
\begin{eqnarray}
   \Gamma = \frac{c_P}{6ac}\frac{\kappa\rho^2 v_c\lambda}{T^3}.
\end{eqnarray}
Using this, the values of $\nabla$ and $\nabla'$ can be determined from
\begin{eqnarray}
      \nabla_r = \nabla - \frac{9}{4}\Gamma(\nabla - \nabla'),\qquad
      \frac{\nabla_r-\nabla}{\nabla_r-\nabla_a} =
      \frac{9\Gamma^2/4}{1+\Gamma(1+9\Gamma/4)},
\end{eqnarray}
where $\nabla_r$ is the radiative temperature gradient. All of these are standard
results of MLT (cf. \citealt{CoxGiuli1968}), but we have taken care here to only
use expressions that do not assume a steady state value for $v_c$ in order to have a
self-consistent model. Although this model incorporates the timescales relevant
to the process, it does not intend to solve some of the long-standing problems
with MLT (see \citealt{Arnett+2018a} for a recent discussion). For
instance, our model does not incorporate overshooting directly but instead
uses an exponentially decaying mixing coefficient beyond convective regions (see
Section \ref{sec:methods}) which does not account for energy transport.
Also, sharp composition gradients near convective
boundaries can lead to discontinuities in the Brunt-V\"ais\"al\"a
frequency, producing a discontinuous $\partial v_c/\partial t$ and $v_c$ at a convective
boundary. Under these circumstances we would expect turbulent energy to be
transported through the boundary, but our model does not include this effect.

\section{Resolution and nuclear reaction network convergence
test}\label{app:conv}
In order to test if our results are converged, we have performed a test using
the first pulse of our $84M_\odot$ model. Using our default
setup, at the onset of the pulse this star has $58.1M_\odot$ and after the first mass
ejection ends up with a mass of $41.49M_\odot$. During this phase, the model is
resolved using between $\sim 2500-3500$ cells and $\sim 6000$ timesteps. To test
the convergence of our model to changes in spatial and temporal resolution, we
have computed a model that after helium depletion approximately doubles both.

\begin{figure}
   \begin{center}
   \includegraphics[width=\columnwidth]{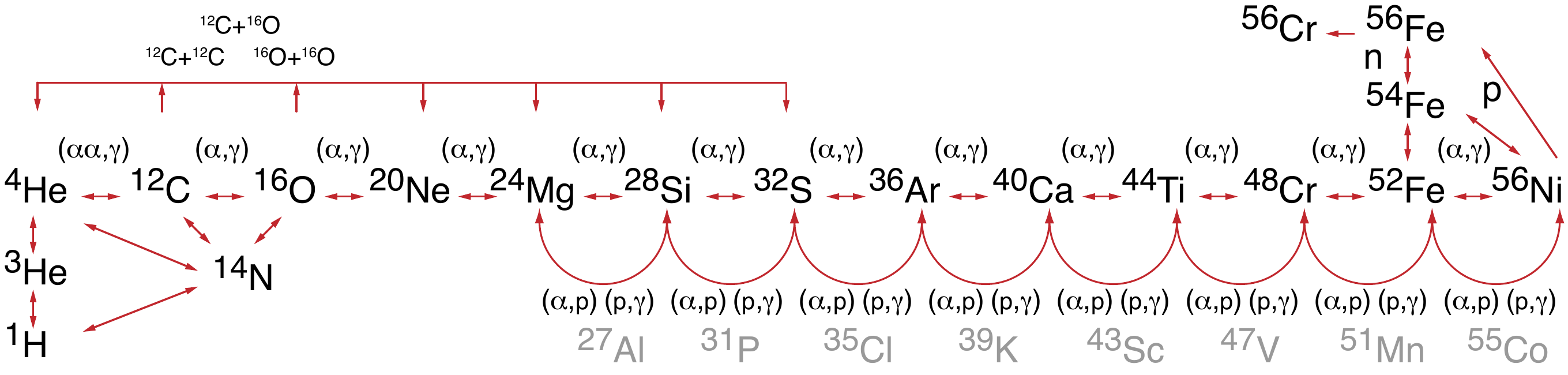}
   \end{center}
   \caption{List of isotopes and linkages in the approx21 network used during
   late burning stages in our calculations.\label{fig:approx21}}
\end{figure}

During the pulsational phase we use the approx21 reaction network  for which the
isotopes and linkages are shown in Figure \ref{fig:approx21}.
The backbone is a strict $\alpha$-chain composed of
($\alpha$,$\gamma$) and ($\gamma$,$\alpha$) links among the 13
isotopes $^4$He, $^{12}$C, $^{16}$O, $^{20}$Ne, $^{24}$Mg, $^{28}$Si,
$^{32}$S, $^{36}$Ar, $^{40}$Ca, $^{44}$Ti, $^{48}$Cr, $^{52}$Fe, and
$^{56}$Ni.  Above $\simeq$ 2.5$\times$10$^{9}$ K is it essential to
include ($\alpha$,p)(p,$\gamma$) and ($\gamma$,p)(p,$\alpha$) links in
order to obtain reasonably accurate energy generation rates and
abundances \citep{Timmes+2000}.  At these elevated
temperatures the flows through the ($\alpha$,p)(p,$\gamma$) sequences
are faster than the flows through the ($\alpha$,$\gamma$) channels.
An ($\alpha$,p)(p,$\gamma$) sequence is, effectively, an
($\alpha$,$\gamma$) reaction through an intermediate isotope.  Approx21
includes 8 ($\alpha$,p)(p,$\gamma$) sequences and their inverses by
assuming steady-state proton flows through the intermediate isotopes
$^{27}$Al, $^{31}$P, $^{35}$Cl, $^{39}$K, $^{43}$Sc, $^{47}$V,
$^{51}$Mn, and $^{55}$Co . The assumed steady-state proton flows
allows inclusion of the ($\alpha$,p)(p,$\gamma$) sequences without
explicitly evolving the proton or intermediate isotope abundances.
In addition to this $\alpha$-chain backbone, approx21 includes
approximations for steady-state hydrogen burning (PP chain and CNO cycle),
carbon and oxygen burning ($^{12}$C+$^{12}$C, $^{12}$C+$^{16}$O, $^{16}$O+$^{16}$O),
and aspects of photodisintegration with $^{54}$Fe. These additions are briefly
described in \citet{Weaver+1978}. Finally, approx21 adds the
$^{56}$Cr and $^{56}$Fe isotopes and tuned steady-state reaction sequences
to attain a reasonably accurate lower electron fraction $Y_e$ (as compared to much larger
reaction networks) for presupernova models \citep{Paxton+2015}. To test the
accuracy of this few-isotope network during a pulse we have also computed the first pulse of
our $84M_\odot$ model using the 203 isotope network of \citet{Renzo+2017} which
is tuned to properly capture silicon burning.

\begin{figure}
   \begin{center}
   \includegraphics[width=\columnwidth]{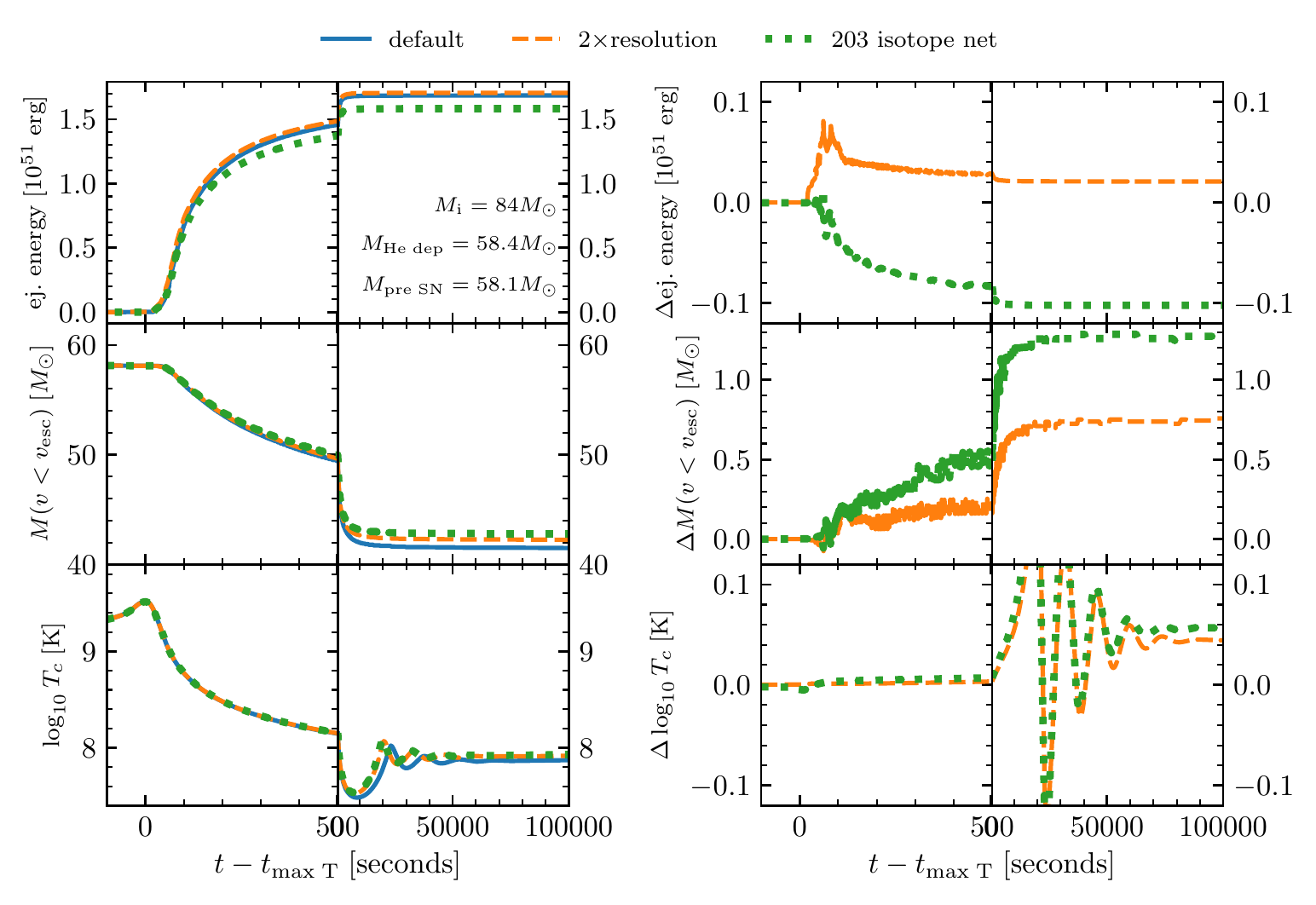}
   \end{center}
   \caption{(left) Evolution of the kinetic energy of ejected layers, the mass at
   velocities below the escape velocity and central temperature
   for the first pulse of an $M_{\rm i}=84M_\odot$ progenitor. Results
   are shown for the default set of parameters used in this paper, a simulation
   with double the resolution in time and space, and one with a 203 isotope
   network rather than the default 21 isotope network we use for all other
   models. (right) For the simulations with higher resolution and a bigger
   network, each line shows how the difference with respect to the simulation
   with our default choice of parameters evolves with time.\label{fig:restest}}
\end{figure}

\begin{figure}
   \begin{center}
   \includegraphics[width=\columnwidth]{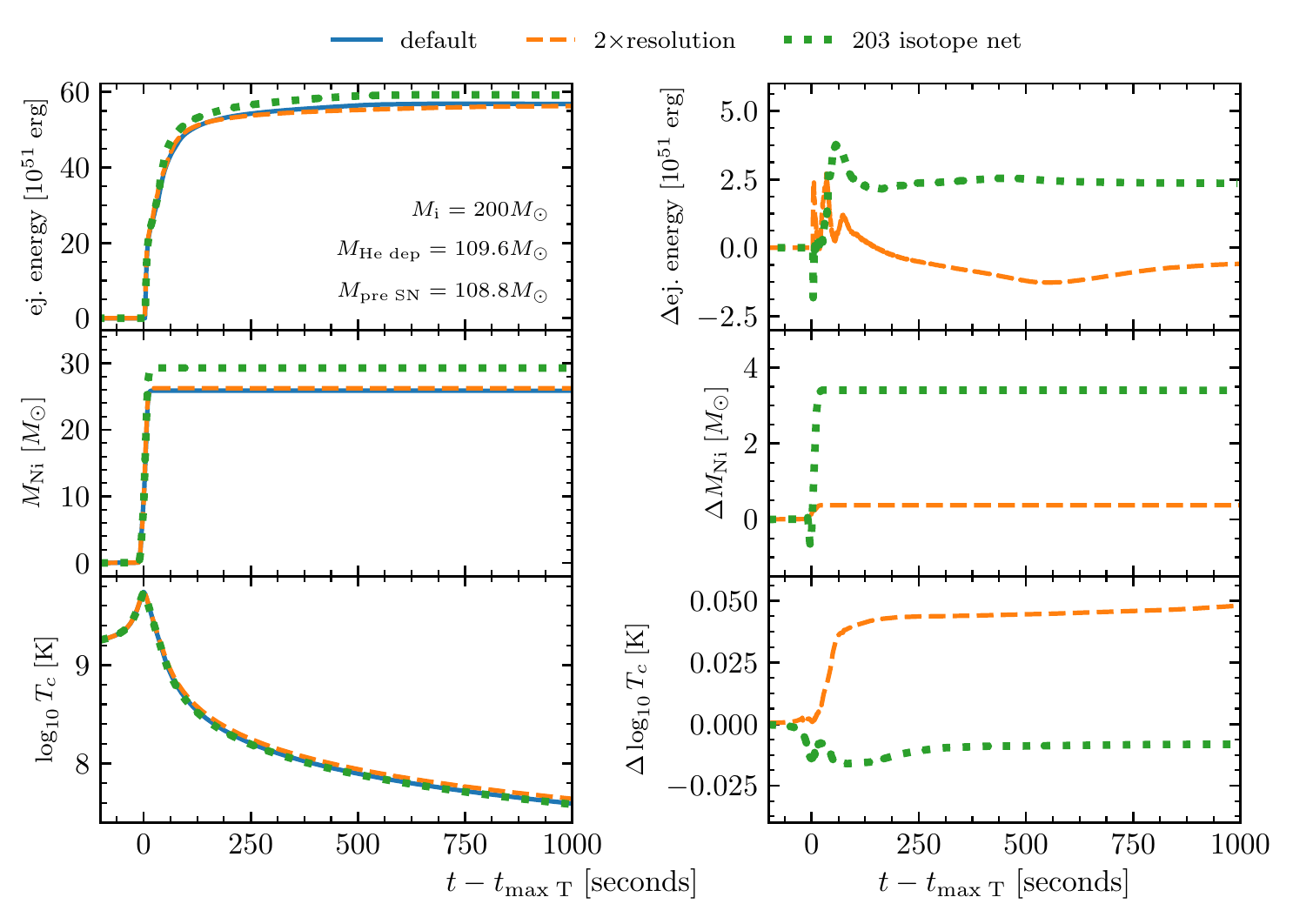}
   \end{center}
   \caption{Same as Figure \ref{fig:restest}, but for a PISN model with
   $M_{i}=200M_\odot$ ($M_{\rm pre\;SN}=108.4M_\odot$), and showing the
   evolution of the total mass of $^{56}$Ni instead of the ejected
   mass.\label{fig:restest2}}
\end{figure}

Figure \ref{fig:restest} shows the results of our convergence tests.
For ease of comparison between
the simulations, we have matched all tracks in time to the point where the first
pulsation reaches its maximum central temperature, and we compare values 100000
seconds after this point.
Overall the three simulations are quantitatively consistent, with relative differences in the kinetic energy
of ejected layers and final masses of around $6\%$. Final central
temperatures digress by around $15\%$, but considering that during the pulse it
is lowered by a factor of $\sim 30$, this is a small error.
Given these results,
and that we do not study detailed nucleosynthetic yields of PPISN or PISN in
this work, we consider our choice of resolution and nuclear reaction network
appropriate. In particular, the use of approx21 instead of the 203 isotope
network reduces the runtime of each model by more than a factor of 10,
significantly lowering the cost of our simulations.

As a more extreme example, we repeat this exercise for a PISN model with an initial mass of
$M_{i}=200M_\odot$, which is shown in Figure \ref{fig:restest2}. This model is near the upper end of the mass range of PISN,
with $M_{\rm pre\;SN}=108.4$, and during the explosion reaches a central
temperature of $5.2\times 10^9\;\rm K$, significantly higher than the first
pulse of the PPISN model shown before, which reaches a maximum temperature of
$3.2\times 10^9\;\rm K$. The model with the approx21 network produces a total of
$25.9M_\odot$ of $^{56}$Ni, which is $\sim 13\%$ lower than that produced by the
model with the 203 isotope network. The error on the kinetic energy of the
ejecta is similar to that of the PPISN model, with the 203 isotope network
predicting an ejecta energy $\sim 4\%$ larger than that of the approx21 model.
We see that the dynamics of the explosion are consistently reproduced, although
the difference in Nickel mass would produce non-negligible differences on the
resulting lightcurves. Our focus however is on the evolution of PPISN, which do
not produce significant amounts of Nickel, so we still consider the use of approx21 to
be justified.

\section{Precision of the relaxation procedure}\label{app:relax}
To model the long-lived phases between pulses in our more massive
progenitors, we use a relaxation procedure that creates a hydrostatic model from
scratch that matches the
mass, entropy and composition profile after the pulse. This method has been described in
Appendix B of \citet{Paxton+2018}
and here we show how well it reproduces the pre-relaxation model.
In order to perform a relaxation after a pulse, we require that
velocities are below $20\;{\rm km\;s^{-1}}$ and no layers are
moving at more than $50\%$ their local sound speed within the inner $99\%$ of
mass that remains bound. To prevent the relaxation happening when these
thresholds are satisfied during minima and maxima of oscillations, we
require these to be satisfied for at least 100 continuous timesteps.
We also require the neutrino and nuclear
burning luminosities to be below $10^{11}L_\odot$ and $10^{10}L_\odot$
respectively, in order to avoid relaxing the model when the core is evolving on
a timescale of $\sim$days.

\begin{figure}
   \includegraphics[width=0.5\columnwidth]{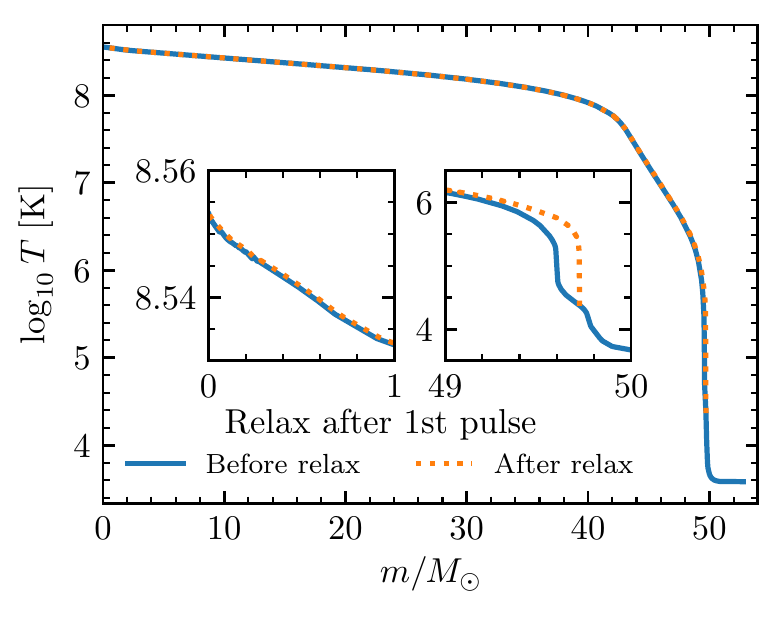}
   \includegraphics[width=0.5\columnwidth]{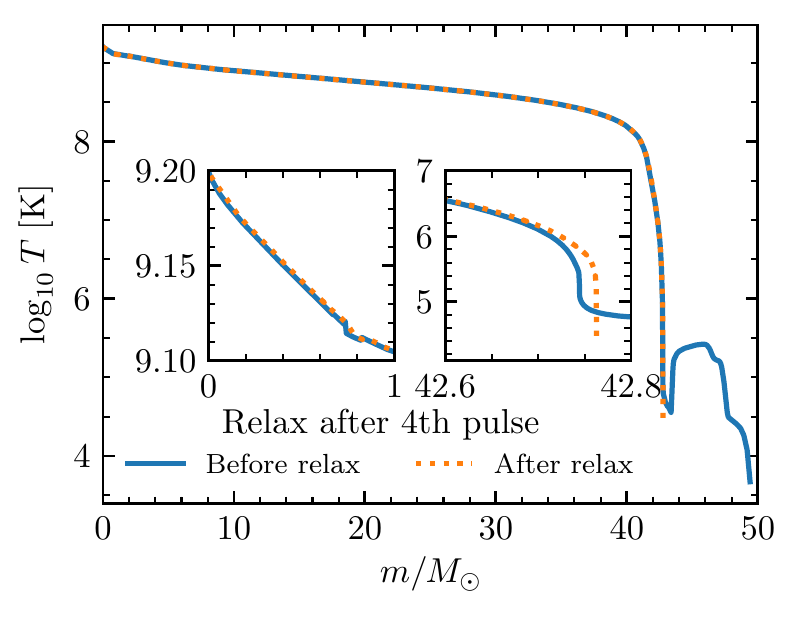}
   \caption{Pre and post-relaxation temperature profiles for our $M_i=76M_\odot$
   model after the first and fourth pulses.\label{fig:relax}}
\end{figure}

Figure \ref{fig:relax} shows the outcome of two relaxation procedures done for the
$76M_\odot$ model shown in Figure \ref{fig:pulses76} after the first and fourth
pulses. For all other three pulses shown, the conditions on the luminosities are
not satisfied, so the model is evolved further without removing the outer
layers. As it can be seen, except for the very outermost layers temperatures are
matched very accurately in the relaxed model, with the central temperature
differing by 0.0002 and 0.0005 dex for the first and fourth pulse respectively.
As expected, the very outermost layers show more noticeable differences, with
clear digressions being visible at the outer $\sim 0.2M_\odot$ and $\sim
0.05M_\odot$ after the first and fourth pulse respectively. Although a
difference is expected, since the very outermost layers are still falling back
when the relaxation is made, we do care about accurately characterizing
observable properties of the star in between pulses. However, the discrepancy
turns out to be not very important. After the first pulse, the thermal timescale of the
outer $0.2M_\odot$ is just of 1.4 years, a very small time
compared to the almost 3 millennia between the first and second pulse. This
means that although we do not trust the effective temperature and luminosity of
our models immediately after a pulse, after $\sim 1$ year any anomalies from
relaxation in the outermost layers will be removed.

\begin{thebibliography}{}
\expandafter\ifx\csname natexlab\endcsname\relax\def\natexlab#1{#1}\fi
\providecommand{\url}[1]{\href{#1}{#1}}
\providecommand{\dodoi}[1]{doi:~\href{http://doi.org/#1}{\nolinkurl{#1}}}
\providecommand{\doeprint}[1]{\href{http://ascl.net/#1}{\nolinkurl{http://ascl.net/#1}}}
\providecommand{\doarXiv}[1]{\href{https://arxiv.org/abs/#1}{\nolinkurl{https://arxiv.org/abs/#1}}}

\bibitem[{{Abbott} {et~al.}(2018{\natexlab{a}}){Abbott}, {Abbott}, {Abbott},
  {Abraham}, \& et~al.}]{GWTC1}
{Abbott}, B.~P., {Abbott}, R., {Abbott}, T.~D., {Abraham}, S., \& et~al.
  2018{\natexlab{a}}, arXiv e-prints, arXiv:1811.12907.
\newblock \doarXiv{1811.12907}

\bibitem[{{Abbott} {et~al.}(2018{\natexlab{b}}){Abbott}, {Abbott}, {Abbott},
  {Abraham}, \& et~al.}]{LIGOpop}
---. 2018{\natexlab{b}}, arXiv e-prints, arXiv:1811.12940.
\newblock \doarXiv{1811.12940}

\bibitem[{{Abbott} {et~al.}(2016){Abbott}, {Abbott}, {Abbott}, {Abernathy},
  {Acernese}, {Ackley}, {Adams}, {Adams}, \& et~al.}]{Abbott_LIGOsummary_2016}
{Abbott}, B.~P., {Abbott}, R., {Abbott}, T.~D., {et~al.} 2016, ArXiv e-prints.
\newblock \doarXiv{1606.04856}

\bibitem[{{Abbott} {et~al.}(2018{\natexlab{c}}){Abbott}, {Abbott}, {Abbott},
  {Abernathy}, {Acernese}, {Ackley}, {Adams}, {Adams}, {Addesso}, {Adhikari},
  \& et~al.}]{Abbott_2018_obsruns}
---. 2018{\natexlab{c}}, Living Reviews in Relativity, 21, 3,
  \dodoi{10.1007/s41114-018-0012-9}

\bibitem[{{Abell} {et~al.}(2009){Abell}, {Allison}, {Anderson}, {Andrew},
  {Angel}, {Armus}, {Arnett}, {Asztalos}, {Axelrod}, \& et~al.}]{Abell+2009}
{Abell}, P.~A., {Allison}, J., {Anderson}, S.~F., {et~al.} 2009, ArXiv
  e-prints.
\newblock \doarXiv{0912.0201}

\bibitem[{{Ade} {et~al.}(2016){Ade}, {Aghanim}, {Arnaud}, {Ashdown}, {Aumont},
  {Baccigalupi}, {Banday}, {Barreiro}, {Bartlett}, \& et~al.}]{PlanckColl2016}
{Ade}, P.~A.~R., {Aghanim}, N., {Arnaud}, M., {et~al.} 2016, \aap, 594, A13,
  \dodoi{10.1051/0004-6361/201525830}

\bibitem[{{Angulo} {et~al.}(1999){Angulo}, {Arnould}, {Rayet}, {Descouvemont},
  {Baye}, {Leclercq-Willain}, {Coc}, {Barhoumi}, {Aguer}, {Rolfs}, {Kunz},
  {Hammer}, {Mayer}, {Paradellis}, {Kossionides}, {Chronidou}, {Spyrou},
  {degl'Innocenti}, {Fiorentini}, {Ricci}, {Zavatarelli}, {Providencia},
  {Wolters}, {Soares}, {Grama}, {Rahighi}, {Shotter}, \& {Lamehi
  Rachti}}]{Angulo+1999}
{Angulo}, C., {Arnould}, M., {Rayet}, M., {et~al.} 1999, Nuclear Physics A,
  656, 3, \dodoi{10.1016/S0375-9474(99)00030-5}

\bibitem[{{Antonini} {et~al.}(2014){Antonini}, {Murray}, \&
  {Mikkola}}]{Antonini+2014}
{Antonini}, F., {Murray}, N., \& {Mikkola}, S. 2014, \apj, 781, 45,
  \dodoi{10.1088/0004-637X/781/1/45}

\bibitem[{{Antonini} {et~al.}(2017){Antonini}, {Toonen}, \&
  {Hamers}}]{Antonini+2017}
{Antonini}, F., {Toonen}, S., \& {Hamers}, A.~S. 2017, \apj, 841, 77,
  \dodoi{10.3847/1538-4357/aa6f5e}

\bibitem[{{Arcavi} {et~al.}(2017){Arcavi}, {Howell}, {Kasen}, {Bildsten},
  {Hosseinzadeh}, {McCully}, {Wong}, {Katz}, {Gal-Yam}, {Sollerman}, {Taddia},
  {Leloudas}, {Fremling}, {Nugent}, {Horesh}, {Mooley}, {Rumsey}, {Cenko},
  {Graham}, {Perley}, {Nakar}, {Shaviv}, {Bromberg}, {Shen}, {Ofek}, {Cao},
  {Wang}, {Huang}, {Rui}, {Zhang}, {Li}, {Li}, {Zhang}, {Valenti}, {Guevel},
  {Shappee}, {Kochanek}, {Holoien}, {Filippenko}, {Fender}, {Nyholm}, {Yaron},
  {Kasliwal}, {Sullivan}, {Blagorodnova}, {Walters}, {Lunnan}, {Khazov},
  {Andreoni}, {Laher}, {Konidaris}, {Wozniak}, \& {Bue}}]{Arcavi+2017}
{Arcavi}, I., {Howell}, D.~A., {Kasen}, D., {et~al.} 2017, \nat, 551, 210,
  \dodoi{10.1038/nature24030}

\bibitem[{{Arnett}(1969)}]{Arnett1969}
{Arnett}, W.~D. 1969, \apss, 5, 180, \dodoi{10.1007/BF00650291}

\bibitem[{{Arnett} {et~al.}(2018){Arnett}, {Meakin}, {Hirschi}, {Cristini},
  {Georgy}, {Campbell}, {Scott}, \& {Kaiser}}]{Arnett+2018a}
{Arnett}, W.~D., {Meakin}, C., {Hirschi}, R., {et~al.} 2018, ArXiv e-prints.
\newblock \doarXiv{1810.04653}

\bibitem[{{Asplund} {et~al.}(2009){Asplund}, {Grevesse}, {Sauval}, \&
  {Scott}}]{Asplund+2009}
{Asplund}, M., {Grevesse}, N., {Sauval}, A.~J., \& {Scott}, P. 2009, \araa, 47,
  481, \dodoi{10.1146/annurev.astro.46.060407.145222}

\bibitem[{{Bartos} {et~al.}(2017){Bartos}, {Kocsis}, {Haiman}, \&
  {M{\'a}rka}}]{Bartos+2017}
{Bartos}, I., {Kocsis}, B., {Haiman}, Z., \& {M{\'a}rka}, S. 2017, \apj, 835,
  165, \dodoi{10.3847/1538-4357/835/2/165}

\bibitem[{{Belczynski} {et~al.}(2014){Belczynski}, {Buonanno}, {Cantiello},
  {Fryer}, {Holz}, {Mandel}, {Miller}, \& {Walczak}}]{Belczynski+2014}
{Belczynski}, K., {Buonanno}, A., {Cantiello}, M., {et~al.} 2014, \apj, 789,
  120, \dodoi{10.1088/0004-637X/789/2/120}

\bibitem[{{Belczynski} {et~al.}(2016{\natexlab{a}}){Belczynski}, {Holz},
  {Bulik}, \& {O'Shaughnessy}}]{Belczynski+2016}
{Belczynski}, K., {Holz}, D.~E., {Bulik}, T., \& {O'Shaughnessy}, R.
  2016{\natexlab{a}}, \nat, 534, 512, \dodoi{10.1038/nature18322}

\bibitem[{{Belczynski} {et~al.}(2016{\natexlab{b}}){Belczynski}, {Heger},
  {Gladysz}, {Ruiter}, {Woosley}, {Wiktorowicz}, {Chen}, {Bulik},
  {O'Shaughnessy}, {Holz}, {Fryer}, \& {Berti}}]{Belczynski+2016b}
{Belczynski}, K., {Heger}, A., {Gladysz}, W., {et~al.} 2016{\natexlab{b}},
  \aap, 594, A97, \dodoi{10.1051/0004-6361/201628980}

\bibitem[{{Bellm}(2014)}]{Bellm2014}
{Bellm}, E. 2014, in The Third Hot-wiring the Transient Universe Workshop, ed.
  P.~R. {Wozniak}, M.~J. {Graham}, A.~A. {Mahabal}, \& R.~{Seaman}, 27--33

\bibitem[{{Blaauw}(1961)}]{Blaauw1961}
{Blaauw}, A. 1961, \bain, 15, 265

\bibitem[{{Boersma}(1961)}]{Boersma1961}
{Boersma}, J. 1961, \bain, 15, 291

\bibitem[{{B{\"o}hm-Vitense}(1958)}]{Bohm-Vitense1958}
{B{\"o}hm-Vitense}, E. 1958, Zeitschrift f\"ur Astrophysik, 46, 108

\bibitem[{{Bond} {et~al.}(1982){Bond}, {Arnett}, \& {Carr}}]{Bond+1982}
{Bond}, J.~R., {Arnett}, W.~D., \& {Carr}, B.~J. 1982, in NATO Advanced Science
  Institutes (ASI) Series C, Vol.~90, NATO Advanced Science Institutes (ASI)
  Series C, ed. M.~J. {Rees} \& R.~J. {Stoneham}, 303--311

\bibitem[{{Breivik} {et~al.}(2016){Breivik}, {Rodriguez}, {Larson}, {Kalogera},
  \& {Rasio}}]{Breivik+2016}
{Breivik}, K., {Rodriguez}, C.~L., {Larson}, S.~L., {Kalogera}, V., \& {Rasio},
  F.~A. 2016, \apjl, 830, L18, \dodoi{10.3847/2041-8205/830/1/L18}

\bibitem[{{Brott} {et~al.}(2011){Brott}, {de Mink}, {Cantiello}, {Langer}, {de
  Koter}, {Evans}, {Hunter}, {Trundle}, \& {Vink}}]{Brott+2011}
{Brott}, I., {de Mink}, S.~E., {Cantiello}, M., {et~al.} 2011, \aap, 530, A115,
  \dodoi{10.1051/0004-6361/201016113}

\bibitem[{{Caughlan} \& {Fowler}(1988)}]{CaughlanFowler1988}
{Caughlan}, G.~R., \& {Fowler}, W.~A. 1988, Atomic Data and Nuclear Data
  Tables, 40, 283, \dodoi{10.1016/0092-640X(88)90009-5}

\bibitem[{{Chan} {et~al.}(2018){Chan}, {M{\"u}ller}, {Heger}, {Pakmor}, \&
  {Springel}}]{Chan+2018}
{Chan}, C., {M{\"u}ller}, B., {Heger}, A., {Pakmor}, R., \& {Springel}, V.
  2018, \apjl, 852, L19, \dodoi{10.3847/2041-8213/aaa28c}

\bibitem[{{Chatterjee} {et~al.}(2017){Chatterjee}, {Rodriguez}, \&
  {Rasio}}]{Chatterjee+2017}
{Chatterjee}, S., {Rodriguez}, C.~L., \& {Rasio}, F.~A. 2017, \apj, 834, 68,
  \dodoi{10.3847/1538-4357/834/1/68}

\bibitem[{{Chatzopoulos} \& {Wheeler}(2012)}]{ChatzopoulosWheeler2012}
{Chatzopoulos}, E., \& {Wheeler}, J.~C. 2012, \apj, 748, 42,
  \dodoi{10.1088/0004-637X/748/1/42}

\bibitem[{{Chatzopoulos} {et~al.}(2013){Chatzopoulos}, {Wheeler}, \&
  {Couch}}]{Chatzopoulos+2013}
{Chatzopoulos}, E., {Wheeler}, J.~C., \& {Couch}, S.~M. 2013, \apj, 776, 129,
  \dodoi{10.1088/0004-637X/776/2/129}

\bibitem[{{Chen} {et~al.}(2014){Chen}, {Woosley}, {Heger}, {Almgren}, \&
  {Whalen}}]{Chen+2014}
{Chen}, K.-J., {Woosley}, S., {Heger}, A., {Almgren}, A., \& {Whalen}, D.~J.
  2014, \apj, 792, 28, \dodoi{10.1088/0004-637X/792/1/28}

\bibitem[{{Clark} {et~al.}(1979){Clark}, {van den Heuvel}, \&
  {Sutantyo}}]{Clark1979}
{Clark}, J.~P.~A., {van den Heuvel}, E.~P.~J., \& {Sutantyo}, W. 1979, \aap,
  72, 120

\bibitem[{{Cox} \& {Giuli}(1968)}]{CoxGiuli1968}
{Cox}, J.~P., \& {Giuli}, R.~T. 1968, {Principles of stellar structure }
  (Gordon \& Breach)

\bibitem[{{de Mink} \& {Mandel}(2016)}]{deMinkMandel2016}
{de Mink}, S.~E., \& {Mandel}, I. 2016, \mnras, 460, 3545,
  \dodoi{10.1093/mnras/stw1219}

\bibitem[{{Dominik} {et~al.}(2012){Dominik}, {Belczynski}, {Fryer}, {Holz},
  {Berti}, {Bulik}, {Mandel}, \& {O'Shaughnessy}}]{Dominik+2012}
{Dominik}, M., {Belczynski}, K., {Fryer}, C., {et~al.} 2012, \apj, 759, 52,
  \dodoi{10.1088/0004-637X/759/1/52}

\bibitem[{{Fishbach} \& {Holz}(2017)}]{FishbachHolz2017}
{Fishbach}, M., \& {Holz}, D.~E. 2017, \apjl, 851, L25,
  \dodoi{10.3847/2041-8213/aa9bf6}

\bibitem[{{Fowler} \& {Hoyle}(1964)}]{FowlerHoyle1964}
{Fowler}, W.~A., \& {Hoyle}, F. 1964, \apjs, 9, 201, \dodoi{10.1086/190103}

\bibitem[{{Fraley}(1968)}]{Fraley1968}
{Fraley}, G.~S. 1968, \apss, 2, 96, \dodoi{10.1007/BF00651498}

\bibitem[{{Fryer}(1999)}]{Fryer1999}
{Fryer}, C.~L. 1999, \apj, 522, 413, \dodoi{10.1086/307647}

\bibitem[{{Gal-Yam} {et~al.}(2009){Gal-Yam}, {Mazzali}, {Ofek}, {Nugent},
  {Kulkarni}, {Kasliwal}, {Quimby}, {Filippenko}, {Cenko}, {Chornock},
  {Waldman}, {Kasen}, {Sullivan}, {Beshore}, {Drake}, {Thomas}, {Bloom},
  {Poznanski}, {Miller}, {Foley}, {Silverman}, {Arcavi}, {Ellis}, \&
  {Deng}}]{Gal-Yam+2009}
{Gal-Yam}, A., {Mazzali}, P., {Ofek}, E.~O., {et~al.} 2009, \nat, 462, 624,
  \dodoi{10.1038/nature08579}

\bibitem[{{Glatzel} {et~al.}(1985){Glatzel}, {Fricke}, \& {El
  Eid}}]{Glatzel+1985}
{Glatzel}, W., {Fricke}, K.~J., \& {El Eid}, M.~F. 1985, \aap, 149, 413

\bibitem[{{Hamann} {et~al.}(1995){Hamann}, {Koesterke}, \&
  {Wessolowski}}]{Hamann+1995}
{Hamann}, W.-R., {Koesterke}, L., \& {Wessolowski}, U. 1995, \aap, 299, 151

\bibitem[{{Hannam} {et~al.}(2013){Hannam}, {Brown}, {Fairhurst}, {Fryer}, \&
  {Harry}}]{Hannam+2013}
{Hannam}, M., {Brown}, D.~A., {Fairhurst}, S., {Fryer}, C.~L., \& {Harry},
  I.~W. 2013, \apjl, 766, L14, \dodoi{10.1088/2041-8205/766/1/L14}

\bibitem[{{Heger} {et~al.}(2000){Heger}, {Langer}, \& {Woosley}}]{Heger+2000}
{Heger}, A., {Langer}, N., \& {Woosley}, S.~E. 2000, \apj, 528, 368,
  \dodoi{10.1086/308158}

\bibitem[{{Heger} \& {Woosley}(2002)}]{HegerWoosley2002}
{Heger}, A., \& {Woosley}, S.~E. 2002, \apj, 567, 532, \dodoi{10.1086/338487}

\bibitem[{{Heger} {et~al.}(2005){Heger}, {Woosley}, \& {Spruit}}]{Heger+2005}
{Heger}, A., {Woosley}, S.~E., \& {Spruit}, H.~C. 2005, \apj, 626, 350,
  \dodoi{10.1086/429868}

\bibitem[{{Herwig}(2000)}]{Herwig2000}
{Herwig}, F. 2000, \aap, 360, 952

\bibitem[{{Hulse} \& {Taylor}(1975)}]{HulseTaylor1975}
{Hulse}, R.~A., \& {Taylor}, J.~H. 1975, \apjl, 195, L51,
  \dodoi{10.1086/181708}

\bibitem[{Hunter(2007)}]{hunter_2007_aa}
Hunter, J.~D. 2007, Computing In Science \&amp; Engineering, 9, 90

\bibitem[{{Iglesias} \& {Rogers}(1996)}]{IglesiasRogers1996}
{Iglesias}, C.~A., \& {Rogers}, F.~J. 1996, \apj, 464, 943,
  \dodoi{10.1086/177381}

\bibitem[{{Ivanova} {et~al.}(2013{\natexlab{a}}){Ivanova}, {Justham}, {Avendano
  Nandez}, \& {Lombardi}}]{Ivanova+2013b}
{Ivanova}, N., {Justham}, S., {Avendano Nandez}, J.~L., \& {Lombardi}, J.~C.
  2013{\natexlab{a}}, Science, 339, 433, \dodoi{10.1126/science.1225540}

\bibitem[{{Ivanova} {et~al.}(2013{\natexlab{b}}){Ivanova}, {Justham}, {Chen},
  {De Marco}, {Fryer}, {Gaburov}, {Ge}, {Glebbeek}, {Han}, {Li}, {Lu}, {Marsh},
  {Podsiadlowski}, {Potter}, {Soker}, {Taam}, {Tauris}, {van den Heuvel}, \&
  {Webbink}}]{Ivanova+2013}
{Ivanova}, N., {Justham}, S., {Chen}, X., {et~al.} 2013{\natexlab{b}}, \aapr,
  21, 59, \dodoi{10.1007/s00159-013-0059-2}

\bibitem[{{Kaaret} {et~al.}(2017){Kaaret}, {Feng}, \& {Roberts}}]{Kaaret+2017}
{Kaaret}, P., {Feng}, H., \& {Roberts}, T.~P. 2017, \araa, 55, 303,
  \dodoi{10.1146/annurev-astro-091916-055259}

\bibitem[{Kluyver {et~al.}(2016)Kluyver, Ragan-Kelley, P{\'e}rez, Granger,
  Bussonnier, Frederic, Kelley, Hamrick, Grout, Corlay,
  {et~al.}}]{kluyver2016jupyter}
Kluyver, T., Ragan-Kelley, B., P{\'e}rez, F., {et~al.} 2016, in Positioning and
  Power in Academic Publishing: Players, Agents and Agendas: Proceedings of the
  20th International Conference on Electronic Publishing, IOS Press, 87

\bibitem[{{Kulkarni} {et~al.}(1993){Kulkarni}, {Hut}, \&
  {McMillan}}]{Kulkarni+1993}
{Kulkarni}, S.~R., {Hut}, P., \& {McMillan}, S. 1993, \nat, 364, 421,
  \dodoi{10.1038/364421a0}

\bibitem[{{Kuroda} {et~al.}(2018){Kuroda}, {Kotake}, {Takiwaki}, \&
  {Thielemann}}]{Kuroda+2018}
{Kuroda}, T., {Kotake}, K., {Takiwaki}, T., \& {Thielemann}, F.-K. 2018,
  \mnras, 477, L80, \dodoi{10.1093/mnrasl/sly059}

\bibitem[{{Langer} {et~al.}(1983){Langer}, {Fricke}, \&
  {Sugimoto}}]{Langer+1983}
{Langer}, N., {Fricke}, K.~J., \& {Sugimoto}, D. 1983, \aap, 126, 207

\bibitem[{{Lipunov} {et~al.}(1997){Lipunov}, {Postnov}, \&
  {Prokhorov}}]{Lipunov1997}
{Lipunov}, V.~M., {Postnov}, K.~A., \& {Prokhorov}, M.~E. 1997, \na, 2, 43,
  \dodoi{10.1016/S1384-1076(97)00007-9}

\bibitem[{{Lunnan} {et~al.}(2018){Lunnan}, {Fransson}, {Vreeswijk}, {Woosley},
  {Leloudas}, {Perley}, {Quimby}, {Yan}, {Blagorodnova}, {Bue}, {Cenko}, {De
  Cia}, {Cook}, {Fremling}, {Gatkine}, {Gal-Yam}, {Kasliwal}, {Kulkarni},
  {Masci}, {Nugent}, {Nyholm}, {Rubin}, {Suzuki}, \& {Wozniak}}]{Lunnan+2018}
{Lunnan}, R., {Fransson}, C., {Vreeswijk}, P.~M., {et~al.} 2018, Nature
  Astronomy, \dodoi{10.1038/s41550-018-0568-z}

\bibitem[{{Maeder}(1987)}]{Maeder1987}
{Maeder}, A. 1987, \aap, 178, 159

\bibitem[{{Mandel} \& {de Mink}(2016)}]{MandeldeMink2016}
{Mandel}, I., \& {de Mink}, S.~E. 2016, \mnras, 458, 2634,
  \dodoi{10.1093/mnras/stw379}

\bibitem[{{Marchant} {et~al.}(2016){Marchant}, {Langer}, {Podsiadlowski},
  {Tauris}, \& {Moriya}}]{Marchant+2016}
{Marchant}, P., {Langer}, N., {Podsiadlowski}, P., {Tauris}, T.~M., \&
  {Moriya}, T.~J. 2016, \aap, 588, A50, \dodoi{10.1051/0004-6361/201628133}

\bibitem[{{Mokiem} {et~al.}(2007){Mokiem}, {de Koter}, {Vink}, {Puls}, {Evans},
  {Smartt}, {Crowther}, {Herrero}, {Langer}, {Lennon}, {Najarro}, \&
  {Villamariz}}]{Mokiem+2007b}
{Mokiem}, M.~R., {de Koter}, A., {Vink}, J.~S., {et~al.} 2007, \aap, 473, 603,
  \dodoi{10.1051/0004-6361:20077545}

\bibitem[{{Moriya} {et~al.}(2018){Moriya}, {Nicholl}, \&
  {Guillochon}}]{Moriya+2018}
{Moriya}, T.~J., {Nicholl}, M., \& {Guillochon}, J. 2018, ArXiv e-prints.
\newblock \doarXiv{1806.00090}

\bibitem[{{Nieuwenhuijzen} \& {de Jager}(1990)}]{NieuwenhuijzendeJager1990}
{Nieuwenhuijzen}, H., \& {de Jager}, C. 1990, \aap, 231, 134

\bibitem[{{Nishizawa} {et~al.}(2016){Nishizawa}, {Berti}, {Klein}, \&
  {Sesana}}]{Nishizawa+2016}
{Nishizawa}, A., {Berti}, E., {Klein}, A., \& {Sesana}, A. 2016, \prd, 94,
  064020, \dodoi{10.1103/PhysRevD.94.064020}

\bibitem[{{Ott} {et~al.}(2018){Ott}, {Roberts}, {da Silva Schneider}, {Fedrow},
  {Haas}, \& {Schnetter}}]{Ott+2018}
{Ott}, C.~D., {Roberts}, L.~F., {da Silva Schneider}, A., {et~al.} 2018, \apjl,
  855, L3, \dodoi{10.3847/2041-8213/aaa967}

\bibitem[{{Paczynski}(1976)}]{Paczynski1976}
{Paczynski}, B. 1976, in IAU Symposium, Vol.~73, Structure and Evolution of
  Close Binary Systems, ed. P.~{Eggleton}, S.~{Mitton}, \& J.~{Whelan}, 75

\bibitem[{{Paxton} {et~al.}(2011){Paxton}, {Bildsten}, {Dotter}, {Herwig},
  {Lesaffre}, \& {Timmes}}]{Paxton+2011}
{Paxton}, B., {Bildsten}, L., {Dotter}, A., {et~al.} 2011, \apjs, 192, 3,
  \dodoi{10.1088/0067-0049/192/1/3}

\bibitem[{{Paxton} {et~al.}(2013){Paxton}, {Cantiello}, {Arras}, {Bildsten},
  {Brown}, {Dotter}, {Mankovich}, {Montgomery}, {Stello}, {Timmes}, \&
  {Townsend}}]{Paxton+2013}
{Paxton}, B., {Cantiello}, M., {Arras}, P., {et~al.} 2013, \apjs, 208, 4,
  \dodoi{10.1088/0067-0049/208/1/4}

\bibitem[{{Paxton} {et~al.}(2015){Paxton}, {Marchant}, {Schwab}, {Bauer},
  {Bildsten}, {Cantiello}, {Dessart}, {Farmer}, {Hu}, {Langer}, {Townsend},
  {Townsley}, \& {Timmes}}]{Paxton+2015}
{Paxton}, B., {Marchant}, P., {Schwab}, J., {et~al.} 2015, \apjs, 220, 15,
  \dodoi{10.1088/0067-0049/220/1/15}

\bibitem[{{Paxton} {et~al.}(2018){Paxton}, {Schwab}, {Bauer}, {Bildsten},
  {Blinnikov}, {Duffell}, {Farmer}, {Goldberg}, {Marchant}, {Sorokina},
  {Thoul}, {Townsend}, \& {Timmes}}]{Paxton+2018}
{Paxton}, B., {Schwab}, J., {Bauer}, E.~B., {et~al.} 2018, \apjs, 234, 34,
  \dodoi{10.3847/1538-4365/aaa5a8}

\bibitem[{P{\'e}rez \& Granger(2007)}]{perez2007ipython}
P{\'e}rez, F., \& Granger, B.~E. 2007, Computing in Science \& Engineering, 9,
  21

\bibitem[{{Peters}(1964)}]{Peters1964}
{Peters}, P.~C. 1964, Physical Review, 136, 1224,
  \dodoi{10.1103/PhysRev.136.B1224}

\bibitem[{{Portegies Zwart} \& {McMillan}(2000)}]{PortegieszwartMcmillan2000}
{Portegies Zwart}, S.~F., \& {McMillan}, S.~L.~W. 2000, \apjl, 528, L17,
  \dodoi{10.1086/312422}

\bibitem[{{Rakavy} \& {Shaviv}(1967)}]{RakaviShaviv1967}
{Rakavy}, G., \& {Shaviv}, G. 1967, \apj, 148, 803, \dodoi{10.1086/149204}

\bibitem[{{Renzo} {et~al.}(2017){Renzo}, {Ott}, {Shore}, \& {de
  Mink}}]{Renzo+2017}
{Renzo}, M., {Ott}, C.~D., {Shore}, S.~N., \& {de Mink}, S.~E. 2017, \aap, 603,
  A118, \dodoi{10.1051/0004-6361/201730698}

\bibitem[{{Rodriguez} {et~al.}(2016){Rodriguez}, {Chatterjee}, \&
  {Rasio}}]{Rodriguez+2016a}
{Rodriguez}, C.~L., {Chatterjee}, S., \& {Rasio}, F.~A. 2016, \prd, 93, 084029,
  \dodoi{10.1103/PhysRevD.93.084029}

\bibitem[{{Salpeter}(1955)}]{Salpeter1955}
{Salpeter}, E.~E. 1955, \apj, 121, 161, \dodoi{10.1086/145971}

\bibitem[{{Sesana}(2016)}]{Sesana2016}
{Sesana}, A. 2016, Physical Review Letters, 116, 231102,
  \dodoi{10.1103/PhysRevLett.116.231102}

\bibitem[{{Sigurdsson} \& {Hernquist}(1993)}]{SigurdssonHernquist1993}
{Sigurdsson}, S., \& {Hernquist}, L. 1993, \nat, 364, 423,
  \dodoi{10.1038/364423a0}

\bibitem[{{Smith} {et~al.}(2014){Smith}, {Dekany}, {Bebek}, {Bellm}, {Bui},
  {Cromer}, {Gardner}, {Hoff}, {Kaye}, {Kulkarni}, {Lambert}, {Levi}, \&
  {Reiley}}]{Smith+2014}
{Smith}, R.~M., {Dekany}, R.~G., {Bebek}, C., {et~al.} 2014, in Proc. SPIE,
  Vol. 9147, Ground-based and Airborne Instrumentation for Astronomy V, 914779

\bibitem[{Soberman {et~al.}(1997)Soberman, Phinney, \& van~den
  Heuvel}]{Soberman+1997}
Soberman, G.~E., Phinney, E.~S., \& van~den Heuvel, E. P.~J. 1997, \aap, 327,
  620

\bibitem[{{Spera} \& {Mapelli}(2017)}]{SperaMapelli2017}
{Spera}, M., \& {Mapelli}, M. 2017, \mnras, 470, 4739,
  \dodoi{10.1093/mnras/stx1576}

\bibitem[{{Stone} {et~al.}(2017){Stone}, {Metzger}, \& {Haiman}}]{Stone+2017}
{Stone}, N.~C., {Metzger}, B.~D., \& {Haiman}, Z. 2017, \mnras, 464, 946,
  \dodoi{10.1093/mnras/stw2260}

\bibitem[{{Stothers}(1999)}]{Stothers1999}
{Stothers}, R.~B. 1999, \mnras, 305, 365,
  \dodoi{10.1046/j.1365-8711.1999.02444.x}

\bibitem[{{Taam} {et~al.}(1978){Taam}, {Bodenheimer}, \&
  {Ostriker}}]{Taam+1978}
{Taam}, R.~E., {Bodenheimer}, P., \& {Ostriker}, J.~P. 1978, \apj, 222, 269,
  \dodoi{10.1086/156142}

\bibitem[{{Takahashi}(2018)}]{Takahashi2018}
{Takahashi}, K. 2018, \apj, 863, 153, \dodoi{10.3847/1538-4357/aad2d2}

\bibitem[{{Terreran} {et~al.}(2017){Terreran}, {Pumo}, {Chen}, {Moriya},
  {Taddia}, {Dessart}, {Zampieri}, {Smartt}, {Benetti}, {Inserra},
  {Cappellaro}, {Nicholl}, {Fraser}, {Wyrzykowski}, {Udalski}, {Howell},
  {McCully}, {Valenti}, {Dimitriadis}, {Maguire}, {Sullivan}, {Smith}, {Yaron},
  {Young}, {Anderson}, {Della Valle}, {Elias-Rosa}, {Gal-Yam}, {Jerkstrand},
  {Kankare}, {Pastorello}, {Sollerman}, {Turatto}, {Kostrzewa-Rutkowska},
  {Koz{\l}owski}, {Mr{\'o}z}, {Pawlak}, {Pietrukowicz}, {Poleski}, {Skowron},
  {Skowron}, {Soszy{\'n}ski}, {Szyma{\'n}ski}, \& {Ulaczyk}}]{Terreran+2017}
{Terreran}, G., {Pumo}, M.~L., {Chen}, T.-W., {et~al.} 2017, Nature Astronomy,
  1, 713, \dodoi{10.1038/s41550-017-0228-8}

\bibitem[{{Thompson}(2011)}]{Thompson2011}
{Thompson}, T.~A. 2011, \apj, 741, 82, \dodoi{10.1088/0004-637X/741/2/82}

\bibitem[{{Timmes} {et~al.}(2000){Timmes}, {Hoffman}, \&
  {Woosley}}]{Timmes+2000}
{Timmes}, F.~X., {Hoffman}, R.~D., \& {Woosley}, S.~E. 2000, \apjs, 129, 377,
  \dodoi{10.1086/313407}

\bibitem[{{Toro} {et~al.}(1994){Toro}, {Spruce}, \& {Speares}}]{Toro+1994}
{Toro}, E.~F., {Spruce}, M., \& {Speares}, W. 1994, Shock Waves, 4, 25,
  \dodoi{10.1007/BF01414629}

\bibitem[{{Tutukov} \& {Yungelson}(1993)}]{TutukovYungelson1993}
{Tutukov}, A.~V., \& {Yungelson}, L.~R. 1993, \mnras, 260, 675,
  \dodoi{10.1093/mnras/260.3.675}

\bibitem[{{van den Heuvel}(1976)}]{vandenHeuvel1976}
{van den Heuvel}, E.~P.~J. 1976, in IAU Symposium, Vol.~73, Structure and
  Evolution of Close Binary Systems, ed. P.~{Eggleton}, S.~{Mitton}, \&
  J.~{Whelan}, 35

\bibitem[{van~der Walt {et~al.}(2011)van~der Walt, Colbert, \&
  Varoquaux}]{der_walt_2011_aa}
van~der Walt, S., Colbert, S.~C., \& Varoquaux, G. 2011, Computing in Science
  Engineering, 13, 22, \dodoi{10.1109/MCSE.2011.37}

\bibitem[{{Vink} {et~al.}(2001){Vink}, {de Koter}, \& {Lamers}}]{Vink+2001}
{Vink}, J.~S., {de Koter}, A., \& {Lamers}, H.~J.~G.~L.~M. 2001, \aap, 369,
  574, \dodoi{10.1051/0004-6361:20010127}

\bibitem[{{Weaver} {et~al.}(1978){Weaver}, {Zimmerman}, \&
  {Woosley}}]{Weaver+1978}
{Weaver}, T.~A., {Zimmerman}, G.~B., \& {Woosley}, S.~E. 1978, \apj, 225, 1021,
  \dodoi{10.1086/156569}

\bibitem[{{Woosley}(1993)}]{Woosley1993}
{Woosley}, S.~E. 1993, \apj, 405, 273, \dodoi{10.1086/172359}

\bibitem[{{Woosley}(2017)}]{Woosley2017}
---. 2017, \apj, 836, 244, \dodoi{10.3847/1538-4357/836/2/244}

\bibitem[{{Woosley}(2018)}]{Woosley2018}
---. 2018, \apj, 863, 105, \dodoi{10.3847/1538-4357/aad044}

\bibitem[{{Woosley} {et~al.}(2007){Woosley}, {Blinnikov}, \&
  {Heger}}]{Woosley+2007}
{Woosley}, S.~E., {Blinnikov}, S., \& {Heger}, A. 2007, \nat, 450, 390,
  \dodoi{10.1038/nature06333}

\bibitem[{{Woosley} \& {Weaver}(1982)}]{WoosleyWeaver1982}
{Woosley}, S.~E., \& {Weaver}, T.~A. 1982, in NATO Advanced Science Institutes
  (ASI) Series C, Vol.~90, NATO Advanced Science Institutes (ASI) Series C, ed.
  M.~J. {Rees} \& R.~J. {Stoneham}, 79

\bibitem[{{Yoon} {et~al.}(2010){Yoon}, {Woosley}, \& {Langer}}]{Yoon+2010}
{Yoon}, S.-C., {Woosley}, S.~E., \& {Langer}, N. 2010, \apj, 725, 940,
  \dodoi{10.1088/0004-637X/725/1/940}

\bibitem[{{Yoshida} {et~al.}(2016){Yoshida}, {Umeda}, {Maeda}, \&
  {Ishii}}]{Yoshida+2016}
{Yoshida}, T., {Umeda}, H., {Maeda}, K., \& {Ishii}, T. 2016, \mnras, 457, 351,
  \dodoi{10.1093/mnras/stv3002}

\end{thebibliography}

\end{document}